\pdfoutput=1 

\documentclass[
  aps, pre,  twocolumn,  groupedaddress, superscriptaddress 
  showpacs, letterpaper, amsfonts, footinbib, 10pt,  floatfix,
 noeprint, superscriptaddress]{revtex4-2}

\usepackage{hyperref}
\hypersetup{
    colorlinks=true,
    citecolor=blue,
    linkcolor=blue , 
    filecolor=cyan,      
    urlcolor=magenta,
    pdfcreator = {\LaTeX\ and \flqq hyperref\frqq},
}

\RequirePackage{graphicx,amsmath,amssymb,bm}
\usepackage{bbm}
\usepackage{float}
\usepackage{framed} 
\usepackage{amsthm}
\usepackage{caption}
\usepackage{subcaption}
\usepackage[normalem]{ulem}

 \graphicspath{ {./Figs/} } 
 
\newcommand{\bb}{\mathbb}
\newcommand{\tr}{\text{tr}}

\newcommand{\bbz}{\bb{Z}}

\def\be{\begin{equation}}
\def\ee{\end{equation}} 
\def\bsh{\begin{shaded}}
\def\esh{\end{shaded}} 
\def\bpm{\begin{pmatrix}}
\def\epm{\end{pmatrix}}

\begin{document}
\title{Dynamics and Phases of Nonunitary Floquet Transverse-Field Ising Model}
\author{Lei Su}
\affiliation{Department of Physics, University of Chicago, Chicago, Illinois 60637, USA}
\author{Aashish Clerk}
\affiliation{Pritzker School of Molecular Engineering, University of Chicago, Chicago, Illinois 60637, USA}
\author{Ivar Martin}
\affiliation{Materials Science Division, Argonne National Laboratory, Lemont, Illinois 60439, USA}
\affiliation{Department of Physics, University of Chicago, Chicago, Illinois 60637, USA}

\begin{abstract} 
Inspired by current research on measurement-induced quantum phase transitions, we analyze the nonunitary Floquet transverse-field Ising model with complex nearest-neighbor couplings and complex transverse fields. Unlike its unitary counterpart, the model shows a number of steady phases, stable to integrability breaking perturbations. Some phases have robust edge modes and/or spatiotemporal long-range orders in the bulk. The transitions between the phases have extensive entanglement entropy, whose scaling with the system size depends on the number of the real quasiparticle modes in the spectrum at the transition. 
In particular, the volume law scaling appears on some critical lines, protected by pseudo-Hermiticity.
Both the scaling of  entanglement entropy in steady states and the evolution after a quench are compatible with the non-Hermitian generalization of the quasiparticle picture of  Calabrese and Cardy at least qualitatively.
\end{abstract}

\maketitle



\section{Introduction}

Non-equilibrium quantum dynamics in many-body systems is an active area of research cutting across many different subfields of physics.  While nontrivial dynamics can be generated in many different ways, there are two generic routes that  are particularly attractive.  The first is to induce dynamics by a periodic drive, the so-called Floquet approach.  A wealth of recent work shows that many-body Floquet systems can exhibit novel phase transitions \cite{Khemani2016phase, Else2016floquet,Keyserlingk2016absolute,  moessner2017equilibration, harper2020topology}. Realizing nontrivial quantum dynamics in interacting Floquet systems, however,  requires extra ingredients that help suppress or completely avoid heating to an uninteresting infinite-temperature state \cite{D'Alessio2014, Lazarides2014, PONTE2015}. In particular, the heating can be mitigated by introducing disorder which localizes energy via many-body localization (MBL) \cite{Khemani2016phase, moessner2017equilibration}, or by adding external dissipation \cite{Dehghani2014dissipative, Sierant2022dissipativefloquet,  mori2023floquet}.

Another, seemingly distinct route to novel non-equilibrium dynamics is to consider the evolution of a monitored many-body quantum system.  By tuning the measurement rate, and considering an ensemble of quantum trajectories (each corresponding to a particular set of measurement outcomes), one can induce a novel class of phase transitions  \cite{li2018quantum, skinner2019meas, chan2019unitary}. Instead of traditional phase transitions due to symmetry breaking, these  measurement-induced phase transitions (MIPT) do not have conventional order parameters, but instead are witnessed by  entanglement properties and other quantum information-theoretical quantities \cite{potter2022entanglement, fisher2023}.  First experimental evidence of such transitions has recently been reported \cite{noel2022measurement, koh2022experimental}. 

 
\begin{figure*}[t]
\includegraphics[width=0.75\textwidth]{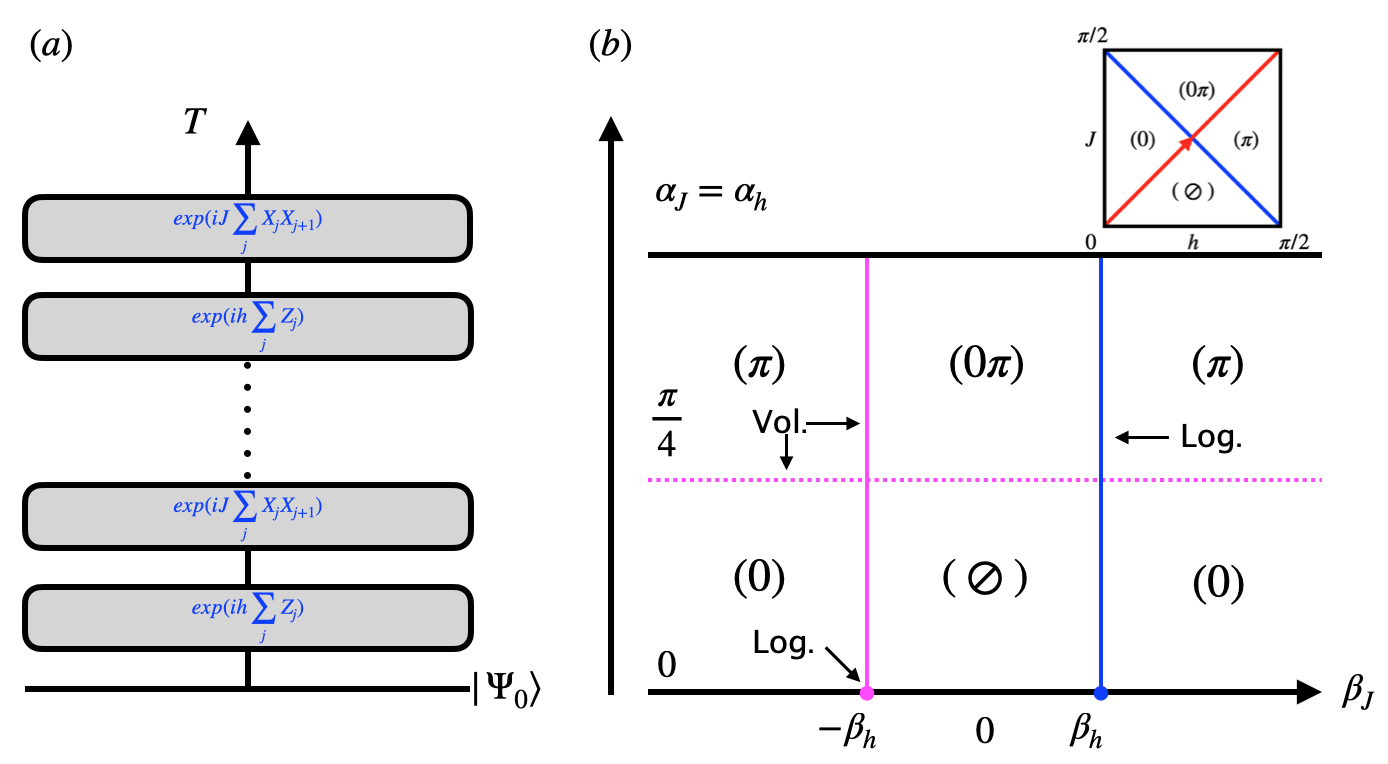}
\centering
\caption{(a) Floquet evolution in Eq.~\ref{floquet}~ of $|\Psi_0\rangle$. (b) Phase diagram of steady states for $ \alpha_J = \alpha_h\equiv \alpha$. $\beta_h >0$ is fixed and $\beta_J$ is varied. It contains four different phases that satisfy an area law in entanglement entropy. The magenta boundaries (solid line, $\alpha_J =\alpha_h = \pi/4$, and dotted line, $\beta_J = -\beta_h$) represent a volume law while the blue solid line represents a logarithmic law. The critical points for $\alpha_J = \alpha_h =0$ have a logarithmic law. The four phases are distinguished by the presence or absence of real $0$- or $\pi$- edge modes in the spectrum with open boundary conditions. In particular, $\oslash$ means no edge modes. Inset: The phase diagram of the unitary Floquet model \cite{Khemani2016phase}; the real parts of the parameters used in the main panel sweep the red diagonal ($\alpha_h = \alpha_J$).}
\label{pd}
\end{figure*}


In this work, we analyze a potentially even richer class of non-equilibrium many-body dynamics realized by combining time-periodic Floquet driving with effective non-unitary evolution (which can be associated with monitoring of the quantum system). Our starting point is the archetypal Floquet transverse-field Ising model (TFIM), where a one-dimensional (1D) lattice of spins (described by Pauli operators $X_j,Y_j,Z_j$) evolves in each drive period according to the operator:  
\begin{align}
    U_F = e^{i J \sum_j X_j X_{j+1}} e^{i h \sum_j Z_j}.
    \label{floquet}
\end{align}
In the well-studied measurement-free case, $J$ and $h$ are real, and correspond respectively to uniform nearest-neighbor Ising interactions and transverse fields. For uniform $J$ and $h$, if the initial state $|\Psi_0\rangle$ (Fig.\ref{pd}(a)) is not a Floquet eigenstate, the system, in the long time limit, may locally approach a periodic version of a generalized Gibbs ensemble \cite{Lazarides2014periodic}. 
On the other hand, if disorder is included, the model exhibits a number of unique non-equilibrium phases  \cite{Khemani2016phase,Bauer2019top, moessner2017equilibration}.   

Instead of the unitary case specified above, here we  consider what happens when the evolution operator $U_F$ is made non-unitary by allowing both $J$ and $h$ to be complex.  This is equivalent  to adding measurement and post-selection to our Floquet dynamics.  To be concrete,  suppose that each $Z_j$ and each bond variable $X_j X_{j+1}$ are continuously monitored using a click style measurement \cite{Turkeshi2021measurement}. For each site $j$ in the lattice there are two click detectors $A_i$ and $B_i$, with the probability of detector $A$
generating a click in some time interval $dt$ being controlled by the operator 
$\hat{A}_i = (1 \pm Z_i)/2$, and the probability of detector $B$ clicking being controlled by the operator $\hat{B}_i = (1 \pm X_j X_{j+1})/2$.  
We specifically consider postselected evolution on experimental runs where no clicks in any of the detectors are recorded.  Using the standard theory of continuously monitored systems (see e.g. Ref.~\cite{wiseman2009quantum}), one can show that the resulting evolution is controlled by a non-Hermitian Hamiltonian. The anti-Hermitian part is generated  by the imaginary parts of $J$ and $h$, which we denote $\beta_J$ and $ \beta_h$; they  correspond respectively to the strengths of the $A_j$ and $B_j$ measurements. We note that special cases and particular aspects have been discussed in, e.g., Refs.~\cite{basu2022Fisher, Ravindranath2023, granet2023volume}.

The Floquet non-unitary TFIM allows us to study the interplay of Floquet driving and measurement-induced dynamics in the simpler setting where the evolution is deterministic (the specific post-selection that we use eliminates the stochasticity inherent in quantum measurement).  Despite the relative simplicity, we find  a number of interesting effects:  
\begin{itemize}
    \item There is a rich variety of steady-state (long-time) phases, which can be characterized by the presence or absence of boundary Majorana modes in the fermionic language when open boundary conditions are used [see Fig.~\ref{pd}(b)].  
    \item The entanglement entropies of the steady states can be understood in terms of the non-Hermitian quasiparticle picture, at least at the spectral level. In particular, non-area-law steady states appear when there are real modes in the spectrum. For example, this occurs on the boundaries between distinct phases, where there is an extensive growth in the steady-state entanglement entropy, with the growth being logarithmic or in some cases volume law.  While some of the volume law behaviors were previously noticed (when the non-uniatry model is a  spacetime-dual to a unitary model \cite{Bertini2018exact, Bertini2019entanglement, Bertini2019exact, Piroli2020exact, Lu2021Spacetime, Ippoliti2022fractal}), we find new volume law regimes that cannot be understood from the duality.  Instead, we tie this new volume-law regime to the pseudo-Hermiticity of the non-Hermitian Floquet Hamiltonian. We show that the topological entanglement entropy that was employed to detect measurement-induced transitions in random quantum circuits \cite{Lavasani2021measurement} can also be used to locate some of these boundaries. This may provide an alternative angle to study Floquet non-Hermitian topological phases (see, e.g., Refs.~\cite{banerjee2023non, zhou2023non}).      
    \item  The postselected measurement-induced dynamics we study allows one to directly stabilize dynamical phases, without the need for disorder, MBL, or additional engineered dissipation. Importantly, this dynamics is robust against (at least) some integrability breaking perturbations. 
    \item Simple conformal field theory (CFT) with complex time can provide  qualitative description of the entanglement entropy evolution, and the ``central charge" of the Floquet criticality is parameter dependent.
\end{itemize}

In the rest of the paper, we substantiate and expand on these results as follows. In Sec.~\ref{sec2}, we describe the spin model and its fermionization as well as formulate the qualitative quasiparticle picture. Furthermore, we analyze the spectrum of the effective Hamiltonian with both periodic boundary conditions (PBCs) and open boundary conditions (OBCs) and study the evolution of entanglement entropy after a quantum quench. We then present the general phase diagram of steady states. In Sec.~\ref{sec3}, we report the detailed numerical analysis of entanglement entropy evolution and scaling, as well as topological entanglement entropy (TEE). We show that the entanglement entropy growth is consistent with the quasiparticle picture at least on the spectral level. Then, in Sec.~\ref{sec4}, we discuss different quench dynamics of an open chain of spins in different phases.  The effect of breaking the integrability is also briefly discussed.  In Sec.~\ref{sec5},  we focus on the $J = h$ case (both complex), and compare the numerical results and the CFT results in both the continuous time limit and the Floquet case.  Finally, in Sec.~\ref{sec6}, we summarize and mention some future directions. Some details are relegated to the Appendix.

\section{Model}
\label{sec2}
In our work, we consider a 1D chain of 1/2-spins undergoing a time-periodic non-unitary evolution $U_F$  described in Eq.~(\ref{floquet}). We start with an initial state $|\Psi_0\rangle$, e.g. a product state and study the quench dynamics [see Fig.~\ref{pd}(a)]. We take $J \equiv \alpha_J +i \beta_J$ and $h \equiv \alpha_h +i \beta_h$ to be complex. Then $U_F$ induces a nonunitary Floquet evolution under non-Hermitian Hamiltonians $H_1 = J \sum_j X_j X_{j+1}$ and $H_2=h \sum_j Z_j$. Since
\be 
U_F = e^{- \beta_J \sum_j X_j X_{j+1}}  e^{i \alpha_J \sum_j X_j X_{j+1}} e^{-\beta_h \sum_j Z_j} e^{i \alpha_h \sum_j Z_j}, 
\ee
$U_F$ can be regarded as a unitary evolution interspersed with imaginary-time evolution. The imaginary time evolution may be achieved by 
introducing couplings to ancillary spins (external to the circuit) that are being measured projectively.
\cite{lin2021real}. 
As discussed in the Introduction, the imaginary-time evolution can also be associated with post-selected measurement-induced dynamics. 

Our study interpolates between and extends beyond several  special cases that have been considered previously in the literature. 
If $J$ and $h$ are real, $U_F$ gives a unitary evolution, and it corresponds to the non-interacting case of the kicked Ising model. Upon making a spacetime duality transformation, i.e. exchanging space and time, $J$ and $h$ become generally complex  and satisfy $\alpha_J=\alpha_h =\pm \pi/4$. There are self dual points at $|J| = |h| = \pi/4$ \cite{Bertini2018exact, Bertini2019entanglement, Bertini2019exact, Piroli2020exact}. At these points, the dual of the original unitary circuit is also unitary, and such a circuit is called dual unitary. Many exact results can be derived using this defining property \cite{Bertini2018exact, Bertini2019entanglement,  Bertini2019exact, Piroli2020exact, Kos2021correlations, claeys2021ergodic}. Away from self dual points, however, the spacetime dual is no longer unitary, and corresponds to a non-Hermitian evolution. 
Naturally, the ``evolution"  along the spatial direction is not independent of the temporal evolution \cite{Ippoliti2022fractal, Ippoliti2021post, Lu2021Spacetime}. For instance, it has been shown that the entanglement entropy scaling (volume vs. area law) of the output of the dual nonunitary circuit is directly related to the  entanglement growth starting from an unentangled initial state in the original time direction (up to some boundary conditions) \cite{Ippoliti2022fractal, bertini2022growth}. It was argued in Ref.~\cite{Lu2021Spacetime} that the impediment to the entaglement growth in the presence of projective measurements is directly related to localization (area law) in the dual circuit. 


When $J$ and $h$ are purely imaginary and small ($\beta_J >0 $ and $\beta_h<0$), we recover continuum limit, which was discussed in Ref.~\cite{kells2021topological}. Our phase diagram includes an area-law-to-area-law phase transition via a logarithmic critical point, similar to the results in Ref.~\cite{Alberton2021Entanglement} (which were for a non-Floquet system).  It was also found there that the phase diagram under the non-Hermitian evolution is smoothly connected to the phase diagram of a continuously monitored  free fermion system. Related ideas have been explored also in Ref.~\cite{Gopalakrishnan2021Entanglement}, where MIPT was studied  in a special model with an effective PT-symmetric  non-Hermitian Hamiltonian. 

Although our model has a large parameter space [spanned by complex $(J,h)$], we will primarily focus here on the interesting case where $\alpha_J = \alpha_h$ (but $\beta_J$ and $\beta_h$ are general).  Physically this corresponds to continuous monitoring of a critical unitary system [red line in the inset of   
Fig.~\ref{pd}(b)].  As we show, the continuous monitoring can drive the system into several distinct  steady states as shown in the main panel of Fig.~\ref{pd}(b).   We also note that while we focus on a particular set of post-selected trajectories (the ``no-click" trajectories), recent work on a related system suggests that the qualitative features here may also be characteristic of all trajectories \cite{kells2021topological}. 


\subsection{The Jordan-Wigner transformation}
To facilitate the analysis, we write the TFIM in terms of complex fermions by using the Jordan-Wigner transformation, 
\be
Z_j = 1- 2c_j^{\dagger} c_j, \quad X_j X_{j+1} = (c_j^{\dagger} -c_j) (c^{\dagger}_{j+1} + c_{j+1}).
\ee
Furthermore, we define the real Majorana modes
\be a_{2j -1} = c_j + c_j^{\dagger}, \quad a_{2j} = i (c_j - c_j^{\dagger}),\ee
which lead to further simplification,
\be
U_F = U_{XX} U_Z = e^{-J \sum_{j =1}^L a_{2j} a_{2j +1}} e^{-h \sum_{j =1}^L a_{2j-1} a_{2j}}.
\ee
Here we have imposed the antiperiodic (periodic) boundary condition on fermions for the even(odd) fermion parity sector. In later discussions, we will also consider the case when OBCs are used.
We can write 
\be H_1 \equiv   i \sum_{j =1}^L a_{2j}  W'_{2j, 2j+1} a_{2j +1}\ee and \be H_2 \equiv   i  \sum_{j =1}^L a_{2j-1} W''_{2j-1, 2j}  a_{2j},\ee
where $W'$ and $W''$ are $2L \times 2L$ matrices. 
Upon application of the Baker-Campbell-Hausdorff (BCH) formula, the bilinear structure of $H_1$ and $H_2$  leads to an effective bilinear Floquet  Hamiltonian  
\be H \equiv    i \sum_{j, k =1}^{2L} a_{j}  W_{j, k} a_{k}
 \label{eqh}
\ee 
with 
\be
 e^{4 W} =e^{ 4 W'} e^{ 4 W''}. 
\ee

 \subsection{The quasiparticle picture}
 \label{secquasiparticle}

The  simple bilinear form of Hamiltonian (\ref{eqh}) implies the existence of non-interacting quasiparticle modes. The quasiparticle picture can be very useful for expressing the time dependence of wavefunctions as well as for interpreting  the entanglement evolution after a quench \cite{Calabrese2005Evolution, alba2017}. Since the effective $H$ is non-Hermitian, the quasiparticles are {\em not} canonical fermions \cite{lee2014heralded, ashida2018full}.

If the non-Hermitian quadratic Hamiltonian $H$ is diagonalizable, then $H = A D A^{-1}$ with diagonal $D$.  The matrix $A$ relates the canonical fermion modes $a_j$ and the quasiparticles $\gamma_{k}$, and
 \be
 H = \sum \epsilon_{k} \tilde{\gamma}_{k}^{\dagger} \gamma_{k}. 
 \ee
Here $\epsilon_{k}$ is in general complex and $ \tilde{\gamma}_{k}^{\dagger} \neq  \gamma_{k}^{\dagger}$. However, the anti-commutation relations   $\{ \tilde{\gamma}_{k}^{\dagger}, \gamma_{k'} \} = \delta_{k, k'}$ are still satisfied. Let us denote $N_{k} \equiv \tilde{\gamma}_{k}^{\dagger} \gamma_{k} $. Then $[N_{k}, N_{k'}] =0$ and $N_{k}^2=N_{k}$. Therefore, we can conveniently use a nonorthonormal basis $|n_{k_1}, n_{k_2}, ..., n_{k_L}\rangle$, such that $N_{k_i}|n_{k_1}, n_{k_2}, ..., n_{k_L}\rangle=n_{k_i}|n_{k_1}, n_{k_2}, ..., n_{k_L}\rangle$ where $n_{k} =0, 1$. The expectation values of $N_{k}$ depend on time, 
\be 
 \langle N_{k}(t)\rangle = \frac{\langle \psi|e^{-i t \sum_{k} \bar{\epsilon}_{k} \gamma_{k}^{\dagger} \tilde{\gamma}_{k}}\ \tilde{\gamma}_{k}^{\dagger} \gamma_{k}\   e^{i t\sum_{k} \epsilon_{k} \tilde{\gamma}_{k}^{\dagger} \gamma_{k}  } |\psi\rangle}{\langle\psi|\psi\rangle}.
\ee
Depending on the sign of  $\text{Im}(\epsilon_{k})$,  $\langle N_{k}\rangle$ will evolve in time to either 0 or 1. In the long time limit, only the real modes with $\text{Im}(\epsilon_{k}) = 0$ play a nontrivial role, similar to the purely real modes in a unitary systems. Such (propagating) modes are expected to play an important role in distributing entanglement through a system starting from unentangled states \cite{Calabrese2005Evolution}.  In the thermodynamic limit, qualitatively, we expect that in the absence of real modes, the final entanglement scaling satisfies an area law.  In contrast, for a finite density of real modes, the entanglement scaling will satisfy a volume law \cite{Calabrese2005Evolution}, whereas if we only have a finite number of real modes the entanglement scaling satisfies the logarithmic law.

There are few points to be noted here. 
First, the system can  be protected by the pseudo-Hermiticity \cite{mostafazadeh2002pseudo} in some parameter regimes. Namely, there exists a Hermitian matrix $\eta$ such that $\eta H \eta^{-1} = H^{\dagger}$. If a pseudo-Hermitian Hamiltonian $H$ is diagonalizable, then it has an antilinear symmetry such as a PT-symmetry. The spectrum of $H$ can be real if the antilinear symmetry is not spontaneously broken. Second, the real eigenmodes and the complex eigenmodes are not necessarily orthogonal, which could affect the quasiparticle picture of entanglement spreading.  Nevertheless, we find that the heuristic non-Hermitian quasiparticle picture described above properly accounts for our numerical results. Similar discussion of the 
 Su-Schrieffer-Heeger model can be found in Ref.~\cite{gal2023} where it was found that the spectrum is dictated by the PT-symmetry and that the entanglement scaling also depends on the (partial) reality of the spectrum  (see also Refs.~\cite{Piccitto2022, Turkeshi2022ent, turkeshi2023entanglement,  zerba2023measurement, paviglianiti2023multipartite}).  

 \subsection{The spectrum}
 \label{spec}
To obtain basic insights into our model, we first consider  the spectrum of the problem. For PBCs, we identify regimes where the system has pseudo-Hermiticity symmetry, which has profound implications for the phase diagram. For PBCs, we find different localized edge Majorana mode configurations that correspond to different phases.
  \subsubsection{Periodic boundary condition: continuous time limit}
Let us first consider the case $|J| =|h|\to 0$; then
\begin{align}
U_F & \approx e^{i J \sum_j X_j X_{j+1}+i h \sum_j Z_j } \nonumber \\ &= e^{i J \sum_j (c_j^{\dagger} -c_j) (c^{\dagger}_{j+1} + c_{j+1}) + i h \sum_j (1- 2c_j^{\dagger} c_j)}.
\label{u_continuous}
\end{align}
Diagonalizing the Hamiltonian in $k$ space gives us the eigenvalues  
\be \lambda_{1,2}=\pm 2\sqrt{h^2 - 2 h J \cos(k) + J^2}.
\label{disp}
\ee
Importantly, it is possible for $\lambda_{1,2}$ to be  real even for complex $J$ and $h$. In those cases, the real modes behave similarly to the Hermitian case and may contribute to a non-area-law behavior in the steady-state entanglement entropy \cite{Calabrese2005Evolution}. We now consider these special cases.

If $J =h$, then $\lambda_{1,2}=\pm 4 J |\sin(k/2)|$. Therefore, for complex $J=h$ with small absolute values,  at $k=0$ we have $\lambda_{1,2}=0$. The quasiparticle picture suggests that the  steady state will have a logarithmic law behavior in entanglement entropy. This is not surprising because all modes but the zero mode will either decay or grow and will not contribute to the entanglement entropy.
 
If $J = \alpha+ i \beta$ and $h =\alpha-i\beta$, then $\lambda_{1,2}=\pm 2\sqrt{2} \sqrt{\alpha^2 -\beta^2 -(\alpha^2 +\beta^2) \cos k}$.  For a given $k$, the modes are either complex conjugates of one another or purely real. The condition $\cos k =(\alpha^2 -\beta^2)/ (\beta^2 +\alpha^2)$ determines the exceptional points in the $k$ space where the eigenvectors coalesce to $(-1, 1)/\sqrt{2}$. 
The existence of real modes and/or complex conjugate pairs is not a coincidence: for these parameters, it is protected by the pseudo-Hermiticity. In fact, if we write the Hamiltonian in Eq.~(\ref{u_continuous}) in the $k$ space, for each $2 \times 2$ $H_k$,  $\eta$ is explicitly given by  
\be
\eta =
\bpm
1 &  \frac{\beta}{\alpha} \cot\left( \frac{k}{2}\right) \\
\frac{\beta}{\alpha} \cot\left( \frac{k}{2}\right) & 1
\epm.  
\ee  
Note that for generic complex $J$ and $h$, there is no PT  or pseudo-Hermiticity symmetry to protect the reality of the spectrum, unlike in Refs.~\cite{Gopalakrishnan2021Entanglement, gal2023}. The pseudo-Hermiticity is important because it allows for the partially real spectrum, with the real modes generating a steady state with a volume law  in entanglement entropy.

\subsubsection{Periodic boundary conditions: general case}
Now let us consider the  Floquet unitary $
U_F$ of Eq.~\ref{floquet} for general parameter values. The spectrum is given by \cite{Lu2021Spacetime}
\be
\varepsilon_k = \pm \sqrt{- w_k^2}, 
\ee
where 
 \be e^{w_k} = \frac{x}{4} \pm \sqrt{\left(\frac{x}{4}\right)^2 -1}
 \ee
with $x =2(1+\cos k) \cos (2h -2 J) + 2 (1-\cos k) \cos (2h +2 J)$. Similar to the continuous-time limit, a real zero mode exists at $k =0$ for $J =h$. It is also easy to check that there are two cases that can lead to an extensive number of real modes:
(1)  $ \alpha_J=  \alpha_h =  \pi/4 \mod \pi/2$ (dual to the unitary case) and (2) $ \alpha_J= \ \alpha_h \neq 0\mod \pi/2, \beta_J =-\beta_h$ (or $ \alpha_J =-\ \alpha_h \neq 0 \mod \pi/2, \beta_J =\beta_h$ in the spacetime dual). 
The spacetime duality is implemented by exchanging time and space coordinates \cite{Bertini2018exact,		 Bertini2019entanglement, Lu2021Spacetime}. In particular,  for $U_F$, the spacetime dual has the form
\be 
\tilde{U}_F = e^{i J' \sum_{j} X_{\tau_j} X_{\tau_{j+1}}} e^{i h' \sum_j Z_{\tau_j}}
\label{Ufp}
\ee
up to some boundary terms. Here 
$J'= -\frac{\pi}{4} - \frac{i}{2} \log \tan h$ and $ h' = -\frac{\pi}{4} - \frac{i}{2} \log \tan J$. For real $J$ and $h$, $U_F$ is unitary, corresponding to the first case.  Indeed, if $J = h =\frac{\pi}{4}  \lambda $, then $J' =h'=- \frac{\pi}{4}+ i \beta$  with $\beta =\frac{1}{2} \log (\tan(\frac{\pi}{4}  \lambda  ))$. It is easy to check that real modes exist between $[0, 2 \pi  \lambda ]$ because $\cos^{-1}(\frac{\cosh (4 \beta) -1}{\cosh (4 \beta) +3}) =2 \pi  \lambda $.  This case is protected by the pseudo-Hermiticity of the effective Hamiltonian  (expressed in terms of complex fermions in the $k$ space)  with the $\eta$ matrix given by 
$\eta =\sigma_x$. The second case is a natural extension of the continuous-time case we discussed in the previous section. It is still protected by pseudo-Hermiticity but  the $\eta$ matrix is more complicated. This extensive number of real modes will produce a volume law in the entanglement entropy in the steady state.

The cases with real modes define the position of the phase boundaries as shown in Fig.~\ref{pd}(b). Indeed, the quasiparticle picture implies that these real modes will not decay or grow and produce a non-area law steady state. We will verify this numerically in the next section.

\begin{figure}[t]
     \centering
     \begin{subfigure}{0.4\textwidth}
         \centering
         \includegraphics[width=1.0\textwidth]{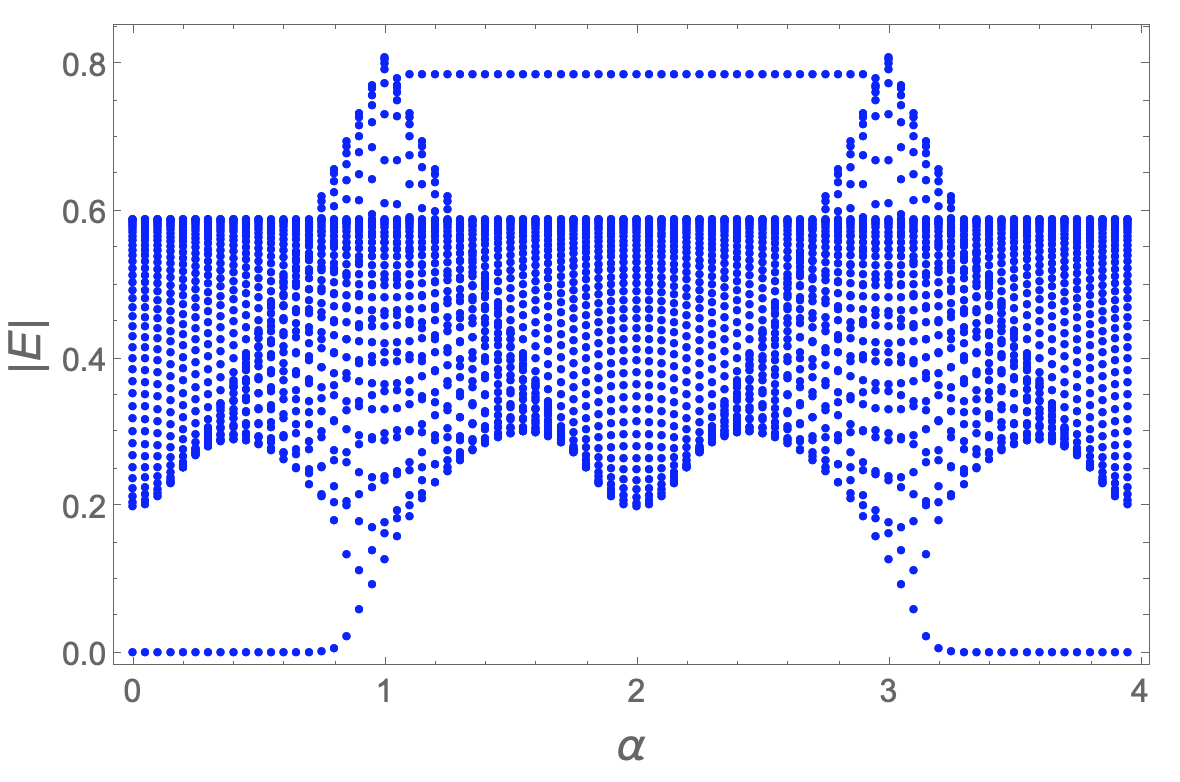}
       
     \end{subfigure}
     \hfill
     \begin{subfigure}{0.4\textwidth}
         \centering
         \includegraphics[width=1.0\textwidth]{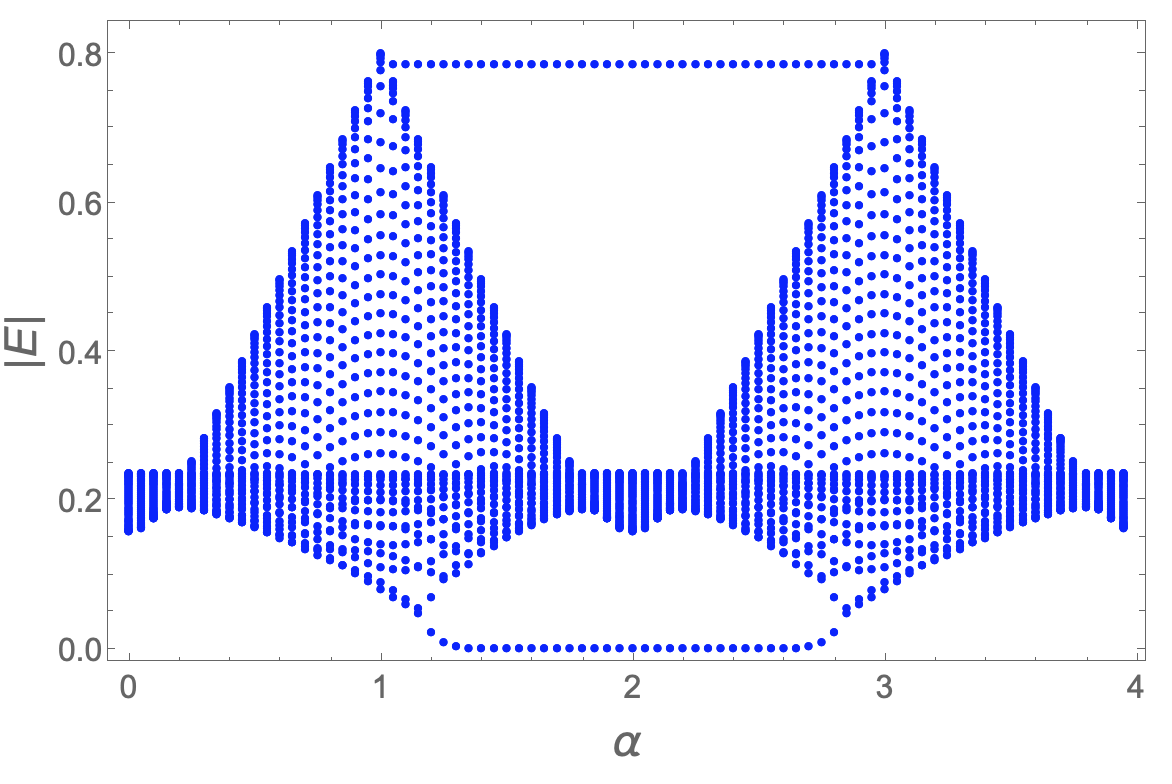}
       
     \end{subfigure} 
       \caption{Absolute values of the spectrum of the Hamiltonian $H$ as a function of $\alpha = \alpha_J = \alpha_h$ when OBC are used. $L=40$.  Top: $\beta_J =-1.0,  \beta_h =0.5  $ (in units of $\pi/4$). Bottom: $\beta_J =-0.1,  \beta_h =0.5  $ (in units of $\pi/4$). } 
       \label{edgemodes}
\end{figure}

\subsubsection{Open boundary condition}
If we use OBCs, it is easy to calculate the spectrum numerically in real space directly. The bulk of the spectrum is not very sensitive to the boundary conditions, but there can also be special edge modes: zero modes and $\pi$ modes. In Fig.~\ref{edgemodes}, we show the absolute value of the spectrum as we tune $\alpha =\alpha_J =\alpha_h$. We can check that these edge-mode energies are real. Because of the reflection symmetry, let us focus on the $0\le \alpha \le \pi/2$ regime. In the upper panel, $\beta_J < -\beta_h$.   When $\alpha \lesssim \pi/4$, the spectrum has zero edge modes, while for  $\alpha \gtrsim \pi/4$, the spectrum has $\pi$ edge modes. In the lower panel, $\beta_J > -\beta_h$.  When $\alpha \lesssim \pi/4$, the spectrum has no edge modes, while when $\alpha \gtrsim \pi/4$, the spectrum has both zero and $\pi$ modes. The presence or absence of the edge modes distinguishes different regimes in  Fig.~\ref{pd}(b) and thus can be used to label the phases. Note that there is a strong finite size effect when $\alpha$ is close to $\pi/4$. As the phase boundaries are approached, the edge modes become non-normalizable for finite $L$.  These edge modes are close relatives of those in the clean unitary case \cite{Thakurathi2013floquet, yates2019almost} as well as the 
Floquet-MBL unitary case \cite{Khemani2016phase}. They are the topological edge modes associated with Floquet symmetry-protected phases \cite{Kitagawa2010top, Potter2016classification, Else2016classification}. They retain real energies even when $J$ and $h$ become complex. We will see later in Sec.~\ref{sec4} that these modes are good indicators of different dynamical behaviors after a quantum quench.

\subsection{Evolution and steady states}
Having established basic spectral features of our model, we now turn to the dynamics in a quench protocol where we start the system in a fermionic Gaussian state and let it evolve under $U_F$. The Floquet Hamiltonian $H$ [Eq.~\ref{eqh}] has the following form in terms of complex fermions:
\be H = \sum \xi_{ij} c_i^{\dagger} c_j + \Delta^-_{ij} c_i c_j  + \Delta^+_{ij} c^{\dagger}_j c^{\dagger}_i.\ee
In general, $ \Delta^+_{ij} \neq  ({\Delta}^-_{ij})^* $.  In terms of real Majorana fermions:
\be H = \sum  H_{ij} a_i a_j ,\ee
and hence 
\be H^{\dagger} = \sum  \bar{H}^T_{ij} a_i a_j.\ee
We want to study the evolution of the states under this generally non-Hermitian Hamiltonian. Since the Hamiltonian is quadratic, a Gaussian state remains Gaussian. We can calculate the entanglement entropy as a function of time from the correlation function $C_{ij}(t) =\langle \psi(t)| a_i a_j|\psi(t) \rangle/\langle \psi(t)|\psi(t) \rangle$, where $|\psi(t)\rangle = \exp(-i H t) |\psi_0\rangle$ and $|\psi_0\rangle$ is the initial state. Under the nonunitary evolution,  
\be
\frac{d C}{dt}  =i  (-C^T \bar{H}^T C+ C^T \bar{H} C +CH C^T - CH^T C^T).  
\ee
Note that since we used the effective Floquet Hamiltonian,  only the correlation functions at multiples of the periods correspond to those in the original system. The time evolution of $C$ can be obtained by solving this equation numerically. If it is a continuous time evolution, the steady state is approached when $d C/dt \rightarrow 0$. In the Floquet setting, the definition of a steady state can be weakened: $C(t +nT) = C(t)$ where $n$ is a positive integer and $T$ is a period also yields a steady state, even if $C(t +T) \ne C(t)$. This case corresponds to discrete time crystals that spontaneously break the discrete time translation symmetry.

If we have a density matrix $\rho$ of a total system comprised of subsystem $A$ and subsystem $B$, we can obtain the reduced density matrix $\rho_A$ by tracing out subsystem $B$: $\rho_A =\tr_B \rho$. Then the von Neumann entanglement entropy is defined as $S_A =- \tr[ \rho_A \ln \rho_A]$. For the free fermion system, it can be evaluated directly using $C'_{ij} \equiv C_{ij} -\delta_{ij}$ \cite{Calabrese2005Evolution}:

\be S_A = -  \tr [\frac{1-C'_A}{2} \ln \frac{1-C'_A}{2} + \frac{1+C'_A}{2} \ln\frac{1+C'_A}{2}],
\label{eqsa}
\ee
where $C'_A$ is $C'_{ij}$ with $i, j$ restricted in $A$. If all eigenvalues of $C'_A$ are denoted as $\{\pm \nu_i \}$, we have \be S_A = - \sum [\frac{1-\nu_i}{2} \ln \frac{1-\nu_i}{2} + \frac{1+\nu_i}{2} \ln\frac{1+\nu_i}{2}].\ee
 
In the case of OBCs, we will also be interested in a generalized TEE, which can be obtained from partitioning the one-dimensional system into four segments as follows:
 \begin{figure}[H]
\includegraphics[width=0.45\textwidth]{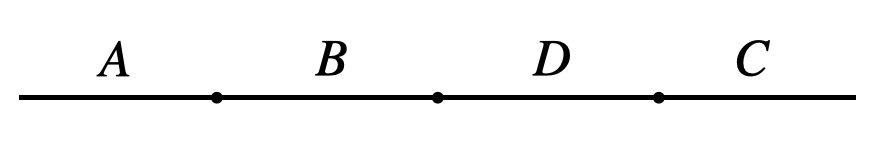}
\centering
\captionsetup{labelformat=empty}  
\end{figure} 
Then TEE is defined as  
\cite{zeng2019quantum, Fromholz2020ent, Lavasani2021measurement, kells2021topological}
 \be 
 S_{\text{top}} = S_{AB} + S_{BC} -S_B -S_{ABC}.
 \label{eqtee}
 \ee 
 This TEE was designed to detect ground-state topological transitions. 
 As we will see in the next section, it can also be used to identify certain transitions in the steady state of our non-unitary Floquet system.

\subsection{Phase diagram}
We now briefly summarize the main features in the  phase diagram of steady states  shown in Fig.~\ref{pd}. The phase diagram can be determined easily by studying the spectra of the Hamiltonian $H$ in Eq.~\ref{eqh}. It depends on the relative magnitude of  $\beta_J$ and $\beta_h$. The phases are demarcated by the lines $|\beta_J|=|\beta_h|$, and $\alpha  = \pi/4$. When $\beta_J = -\beta_h$, the steady-state entanglement entropy satisfies the volume law; when $\beta_J = \beta_h$ the steady state entanglement entropy satisfies the logarithmic law (see next section). The volume-law phase at $\alpha  =\pi/4$ exists because the nonunitary circuit is dual to a unitary circuit, as was already remarked in Ref.~\cite{Lu2021Spacetime}, but also because it is protected by the pseudo-Hermiticity, just as in the case  $\beta_J = -\beta_h$. The rest of the phase diagram has an area law. If  OBCs are imposed, the non-Hermitian spectrum may contain different real-energy edge modes (zero or $\pi$ Majorana modes). Different phases are labeled by the modes present within them: $(\oslash)$ region (no edge modes), $(0)$ region (zero modes), $(\pi)$ region ($\pi$ modes), and $(0 \pi)$ region (both zero and $\pi$ modes).  

In this work, we focused on the $\alpha_J = \alpha_h$ plane on the entire complex manifold where volume law critical lines can be found.  While we have not exhaustively studied parameter regimes $\alpha_J \neq \alpha_h$, a few general comments can be made on differences that emerge in this more general case. If $\alpha_J \neq \alpha_h$, the area-law phases can become logarithmic-law phases depending on $\beta_J$ and $\beta_h$. The transitions between area-law phases become transitions between area-law phases and logarithmic-law phases. These scaling laws are still compatible with the quasiparticle picture at least at the spectral level: if the spectrum of the effective Floquet Hamiltonian contains no real modes, then it has an area-law scaling; on the other hand, if the spectrum contains a few real modes, the scaling is logarithmic. The volume laws we discussed  on critical lines are replaced with logarithmic laws (see, e.g., \cite{granet2023volume}). There is a critical line for $\alpha_J \neq \alpha_h$ that corresponds to the critical line for $\alpha_J = \alpha_h =\pi/4$ while the boundaries at $\beta_J = \pm \beta_h$ remain where they were. 
 Within phases, the edge modes persist, and different quench dynamics with OBC (discussed in Sec.~\ref{sec4}) only weakly depend  on the condition $\alpha_J = \alpha_h$. We leave this detailed discussion to future work.

 \begin{figure}[tb]
\includegraphics[width=0.45\textwidth]{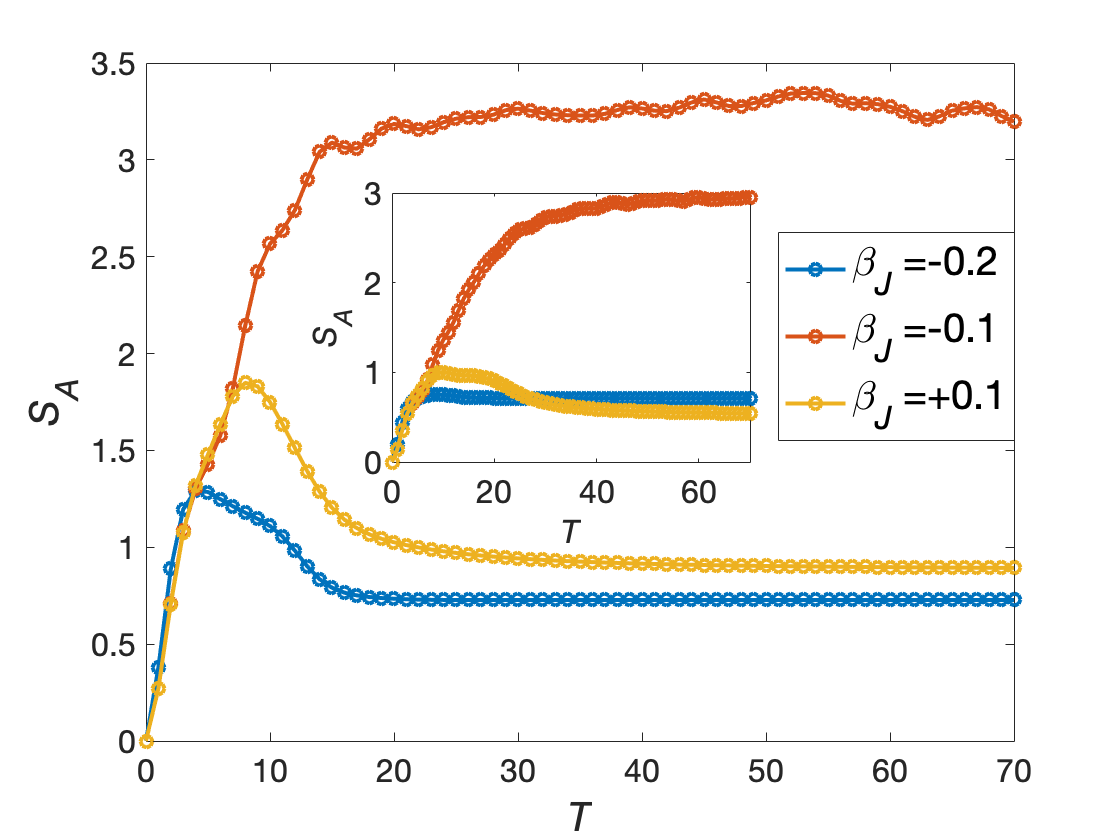}
\centering
\caption{Temporal evolution of entanglement entropy $S_A$ with the PBCs. $\alpha  =0.2 , \beta_h =0.1$ (in units of $\pi/4$). $L_A=8$ and $L_A/L=1/20$. Inset: Evolution of $S_A$  with OBCs. The  subsystem $A$ is chosen to be $[1, L_A]$. If the subsystem $A$ is moved away from the boundary, the curves approach those with PBCs shown in the main panel.}
\label{evolution1}
\end{figure}

\section{Entanglement entropy evolution and scaling}
\label{sec3}
In this section, we present the numerical results on the temporal evolution of entanglement entropy and its scaling in the steady states. For simplicity, we take $L$ to be even and focus on the case in the initial state with odd (even) sites occupied and even (odd) sites empty. 
We will see the entanglement scaling is compatible with the quasiparticle picture for a quench problem. Namely, the steady state can have an area law, a logarithmic law, and a volume law, depending on whether the spectrum of the non-Hermitian Hamiltonian contains no real modes, a few real modes, or an extensive number of real modes, respectively. We also compute the TEE defined in Eq.~\ref{eqtee} and use it to detect the phase transitions at  $\beta_J = \pm \beta_h$.

\subsection{$\alpha  \neq \pi/4$}
\label{alpha1}

\begin{figure}[t]
\centering
\includegraphics[width=0.48\textwidth]{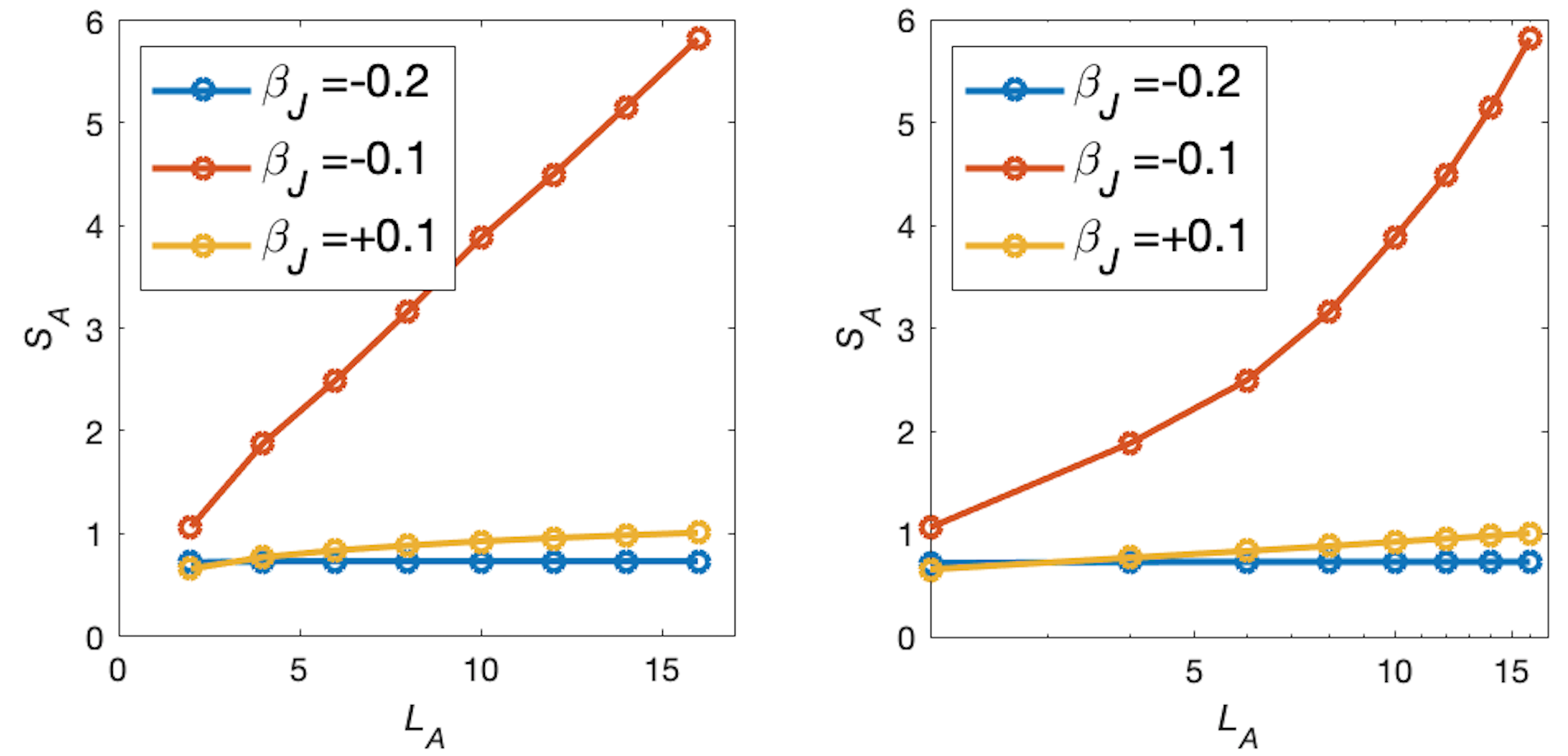}
\caption{Dependence of $S_A$ on the subsystem size $L_A$. Linear (left) and logarithmic (right). $\alpha =0.2 , \beta_h =0.1  $ (in units of $\pi/4$). $L_A/L=1/10$.  }
\label{scaling1}
\end{figure} 

\begin{figure}[b]
\includegraphics[width=0.45\textwidth]{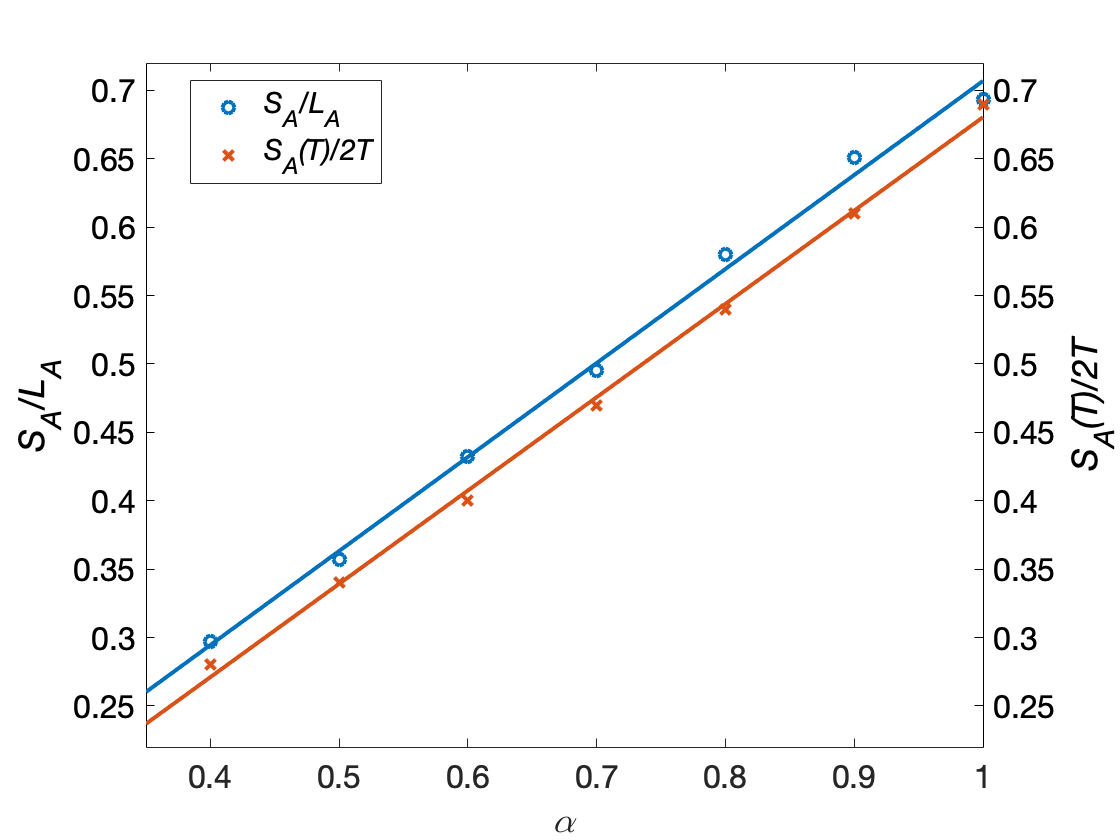}
\centering
\caption{Entanglement entropy density $S_A/L_A$ (blue) and entanglement entropy early growth rate $S_A(T)/2T$ (red) as a function of $\alpha =J = h$ in the spacetime-dual model. $\beta_{J'} =\beta_{h'}= \log (\tan (\alpha))/(\pi/2) $ (in units of $\pi/4$).  $L_A/L=1/10$. Blue and red lines are the respective linear fits.}
\label{den}
\end{figure}

The phase diagram for this case can be seen in Fig.~\ref{pd}(b). We sample three representative points from the phase diagram: $\beta_J =-0.2, -0.1, 0.1$ while fixing $\alpha =0.2$ and $\beta_h =0.1$ (in units of $\pi/4$).  The evolution of the entanglement entropy of subsystem $A$ is shown in Fig.~\ref{evolution1}. The entanglement scaling in different phases is presented in Fig.~\ref{scaling1} both on the linear scale and logarithmic scale.

If the system is in the area-law phase, the entanglement entropy first increases, and possibly drops before it saturates at long times. When $\beta_J = \beta_h$, the qualitative behavior of the evolution curve is similar and the system flows to a steady state with a logarithmic law. This pattern is general and shared by the CFT calculation to be discussed in Sec.~\ref{sec5}. 
When $\beta_J = -\beta_h$, the system approaches a volume law in the long time limit. The evolution curve of $S_A$ is similar to that of a unitary quench: it increases almost linearly at first and then saturates gradually after a time roughly proportional to $L_A$. The wiggly features are due to the finite size effects: increasing the total system size reduces the entanglement revivals and thus smooths the curve. 

In the above discussion, we imposed PBCs. If  different boundary conditions are imposed, e.g., OBCs, the dynamical behavior can be slightly different. With OBCs, the entanglement entropy also depends on the location of subsystem $A$. If $A$ is located deep in the bulk, then the entanglement entropy $S_A$ becomes insensitive to the choice of boundary conditions and the curves approach those with PBCs (see the inset in Fig.~\ref{evolution1}). However, if $A$ sits by the boundary, say, $[1, L_A]$, even though the qualitative features are the same, the saturated values of $S_A$ are slightly different.

\begin{figure}[t]
\includegraphics[width=0.45\textwidth]{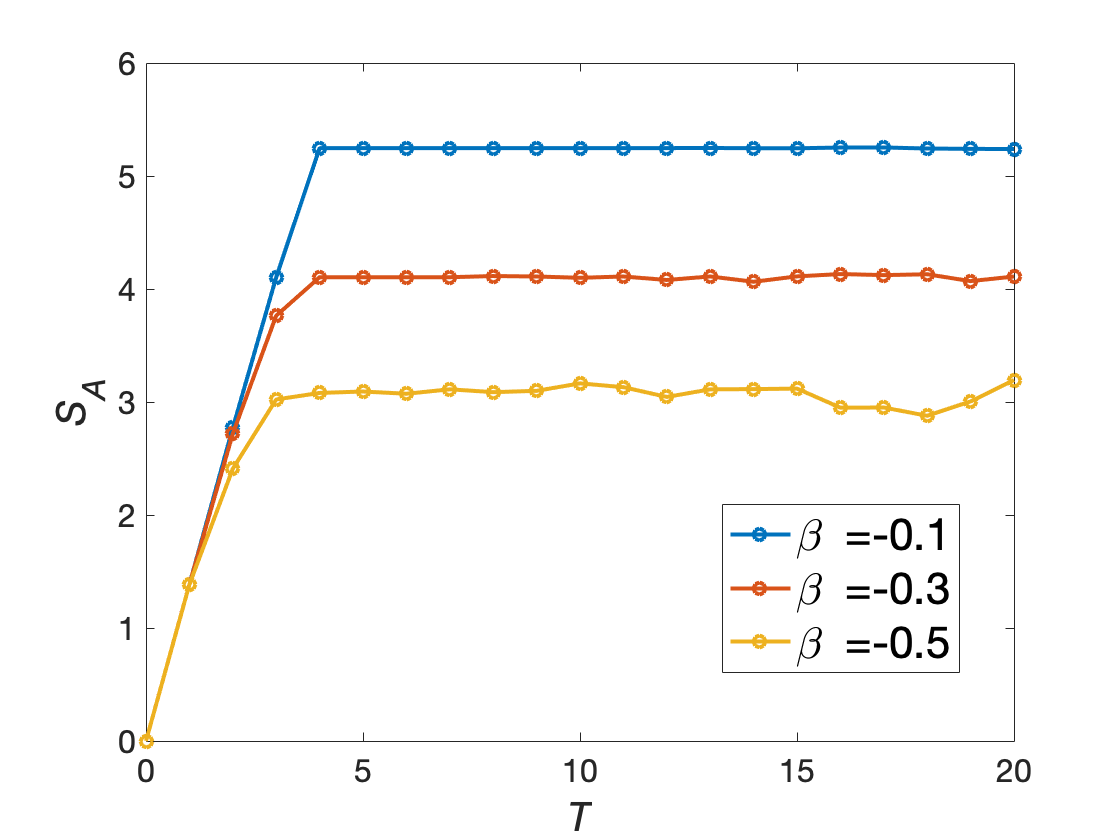}
\centering
\caption{Evolution of $S_A$ for different $\beta =\beta_{J}  =\beta_{h} $ and fixed $\alpha  = 1$ (in units of $\pi/4$).  $L_A=8$, $L_A/L=1/25$. PBCs are used. }
\label{evolution2}
\end{figure}
 
\subsection{$\alpha  = \pi/4$}
When $\alpha  = \pi/4$, the nonunitary circuit is dual to a unitary circuit.  It sits on one of the critical lines of the phase diagram [Fig.~\ref{pd}(b)].  
By resorting to the spacetime duality, it is easy to see that the steady state under the evolution of $U_F$ has a volume-law scaling [dotted magenta line in Fig.~\ref{pd}(b)].  The volume law is compatible with the observation in Ref. \cite{Ippoliti2022fractal} that an area-law steady state is ruled out (up to boundary conditions and fine tuning) for a nonunitary circuit that is dual to a unitary circuit which produces a non-area-law steady state. Previous works found that volume-law entanglement in a free-fermion chain is destroyed in the presence of arbitrarily weak measurements \cite{Cao2019Entanglement, Fidkowski2021how}, and it was suggested in Ref.~\cite{Lu2021Spacetime} that the volume law is symmetry protected by the unitarity of the unrotated circuit. However, as we have already mentioned, more generally, the volume law phase is protected by the pseudo-Hermiticity of the effective Hamiltonian. 

We also mentioned in Sec.~\ref{spec} that if $J = h  =\alpha $ in the unitary circuit, then $J' =h'=- \frac{\pi}{4}+ i \beta$  with $\beta =\frac{1}{2} \log (\tan(\alpha ))$ and the real modes exist in the interval $[0, 8 \alpha]$. In fact, we checked numerically that the entanglement entropy density $S_A/L_A$ is almost linear in $\alpha$ if $\alpha$ is not too small (Fig.~\ref{den}). Plotted together is the initial entanglement entropy growth rate. We see indeed that these two quantities are almost the same, which is not surprising because they are related by the spacetime duality \cite{Ippoliti2022fractal,bertini2022growth} (up to boundary conditions and finite-size and finite-time effects).

We also present the time evolution of $S_A$ of the nonunitary case with different $\beta = \beta_{J'}  =\beta_{h'}$ in Fig.~\ref{evolution2}. We see that $S_A$ increases almost linearly for small $T$ and then saturates. As we increase $\beta$, since more modes become complex, $S_A$ of the steady state decreases. Since $S_A/L_A$ is almost linear in $\alpha$, the dependence of $S_A/L_A$ on $\beta$ can be easily obtained by inverting the relation between $\beta$ and $\alpha$. In particular, if $\alpha$ is close to $\pi/4$, $\beta \approx \alpha-\pi/4 $, and thus $S_A/L_A$ is also linear in $\beta$ for small $\beta$.

\begin{figure}[tb]
\centering
\begin{subfigure}{0.45\textwidth}
\includegraphics[width=1.0\textwidth]{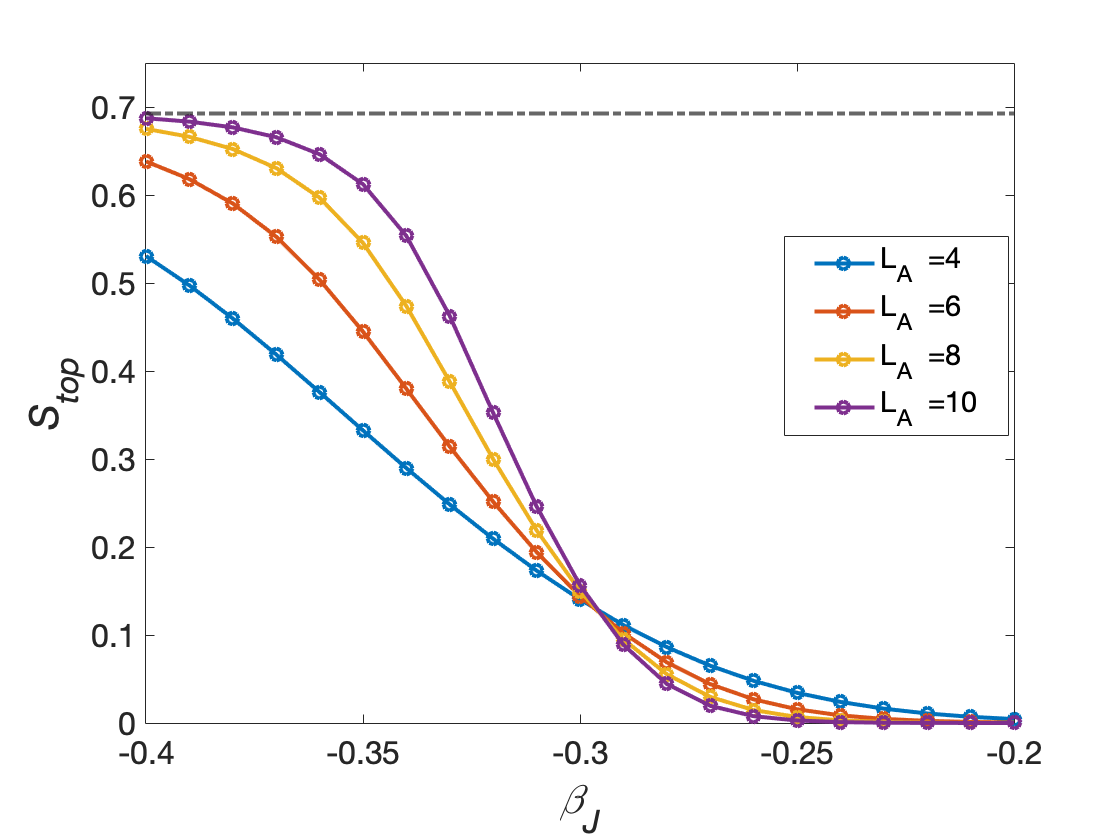}
\centering
\end{subfigure} 
\hfill
\begin{subfigure}{0.45\textwidth}
\includegraphics[width=1.0\textwidth]{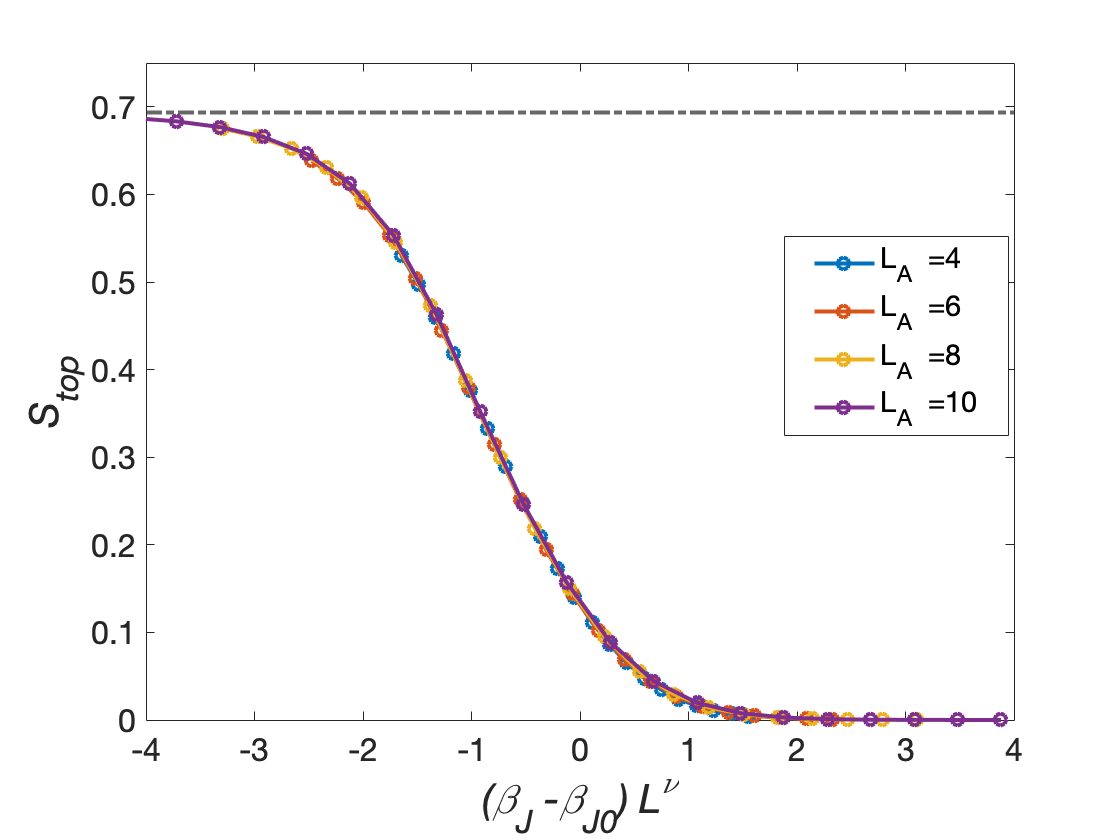}
\centering 
\end{subfigure}
\caption{(a) Topological phase transition between steady states, tuned by $\beta_J$ and detected by TEE. $\alpha  =0.2 , \beta_h =-0.3$ (in units of $\pi/4$). $L_A/L=1/4$. OBCs are used.  The dashed line marks $S_{\text{top}} = \ln 2$.  (b) Scaling collapse of the data, obtained for  $\beta_{J0} \approx \beta_h$ and $\nu \approx 1$.}
\label{finitesize}
\end{figure}

 \begin{figure*}[t]
\includegraphics[width=0.94\textwidth]{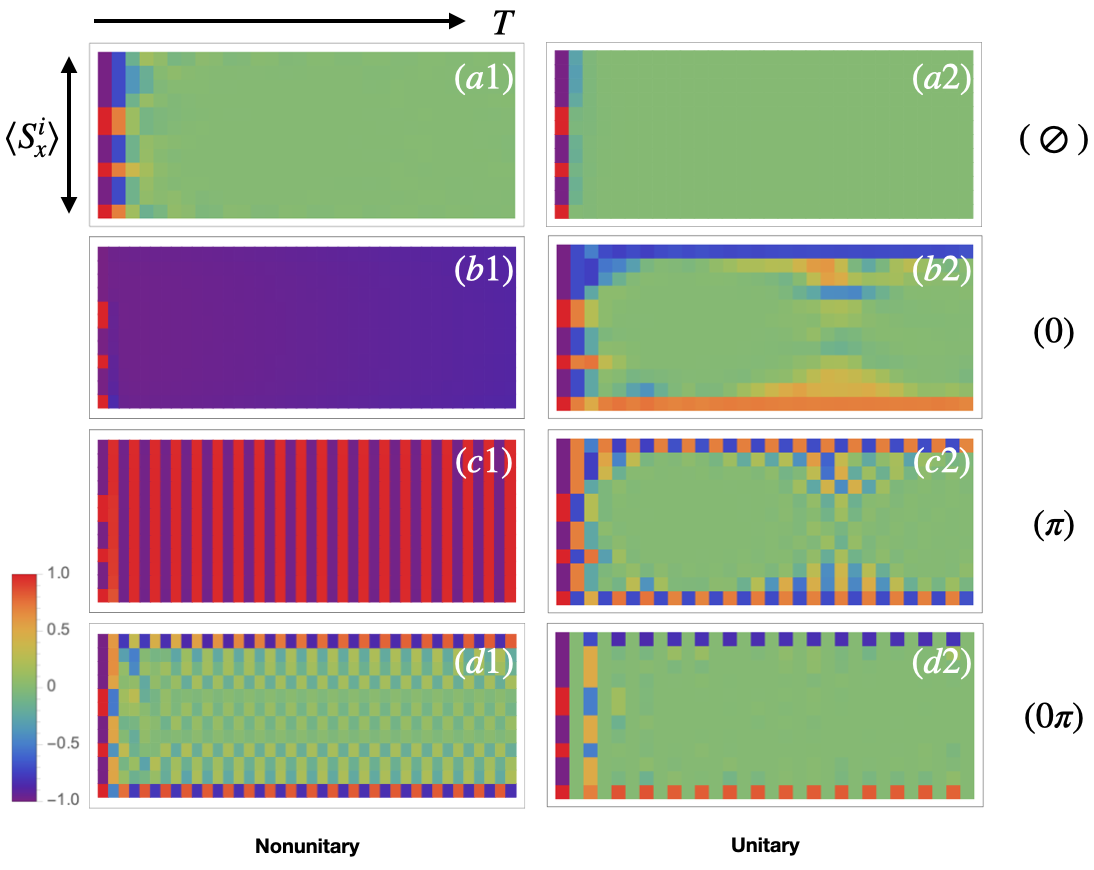}
\centering
\caption{Evolution of $\langle S_x^i\rangle$ in different phases (labeled along the right edge) in the nonunitary (left) and the unitary (right) case. Here, $i =1, 2,..., L$; $L =12$. In each panel, the spatial direction is vertical and the temporal direction is from left to right. The initial states are product states of spins (red: $\mid \uparrow\rangle$; purple: $\mid \downarrow\rangle$). (a1) $\alpha= 0.5$, $\beta_J = - 0.5 , \beta_h= 1.5 $. (b1) $\alpha= 0.5$, $\beta_J = - 1.5 , \beta_h= 0.5 $.   (c1) $\alpha= 1.5$, $\beta_J = - 1.5 , \beta_h= 0.5 $. (d1) $\alpha= 1.5$, $\beta_J =  -0.1 , \beta_h= 0.5 $. (a2) $\alpha_J= 0.5$, $\alpha_h= 1.0$. (b2) $\alpha_J= 1.0$, $\alpha_h= 0.5$. (c2) $\alpha_J= 1.5$, $\alpha_h= 1.0$. (d2) $\alpha_J= 1.5$, $\alpha_h= 1.0$.  All parameters are in units of $\pi/4$.}
\label{phase4}
\end{figure*}

\subsection{Topological entanglement entropy}

We  also compute the TEE [Eq.~(\ref{eqtee})] of different phases and show that it can always be used to detect the transition between $(0)$ phase and the trivial $(\oslash)$ phases. In Fig.~\ref{finitesize}, TEE is shown for fixed $\alpha $ and $\beta_h$, and varying $\beta_J$. The TEE approaches zero for small $|\beta_J|$ and $\ln 2$ when $|\beta_J|$ is large. The curves associated with different sizes cross at a point where $\beta_{J0}  \approx \beta_h$. The contribution to $\ln 2$ comes from a pair of Majorana modes ($ 2 \ln\sqrt{2}$). The finite size scaling is presented in the lower panel by plotting $S_{\text{top}}$ against $(\beta_J -\beta_{J0}) L^{\nu}$. All data collapse almost perfectly onto a single curve. The critical exponent $\nu$ is determined to be approximately 1. 
 
The TEE is also quantized to $\ln 2$ in the $(\pi)$ phase as long as it stays away from the special line $\alpha  = \pi/4$ where a phase transition occurs. We can interpret the $(\pi)$ phase as alternation between two states in the $(0)$ phase with opposite polarization. Then the TEE in the $(\pi)$ phase at one instant is the same as that in the $(0)$ phase. The topological information in the $(0\pi)$-phase is more subtle, and we may need to construct a new quantity to extract it, which we leave for future work.

\begin{figure}[t]
\includegraphics[width=0.49\textwidth]{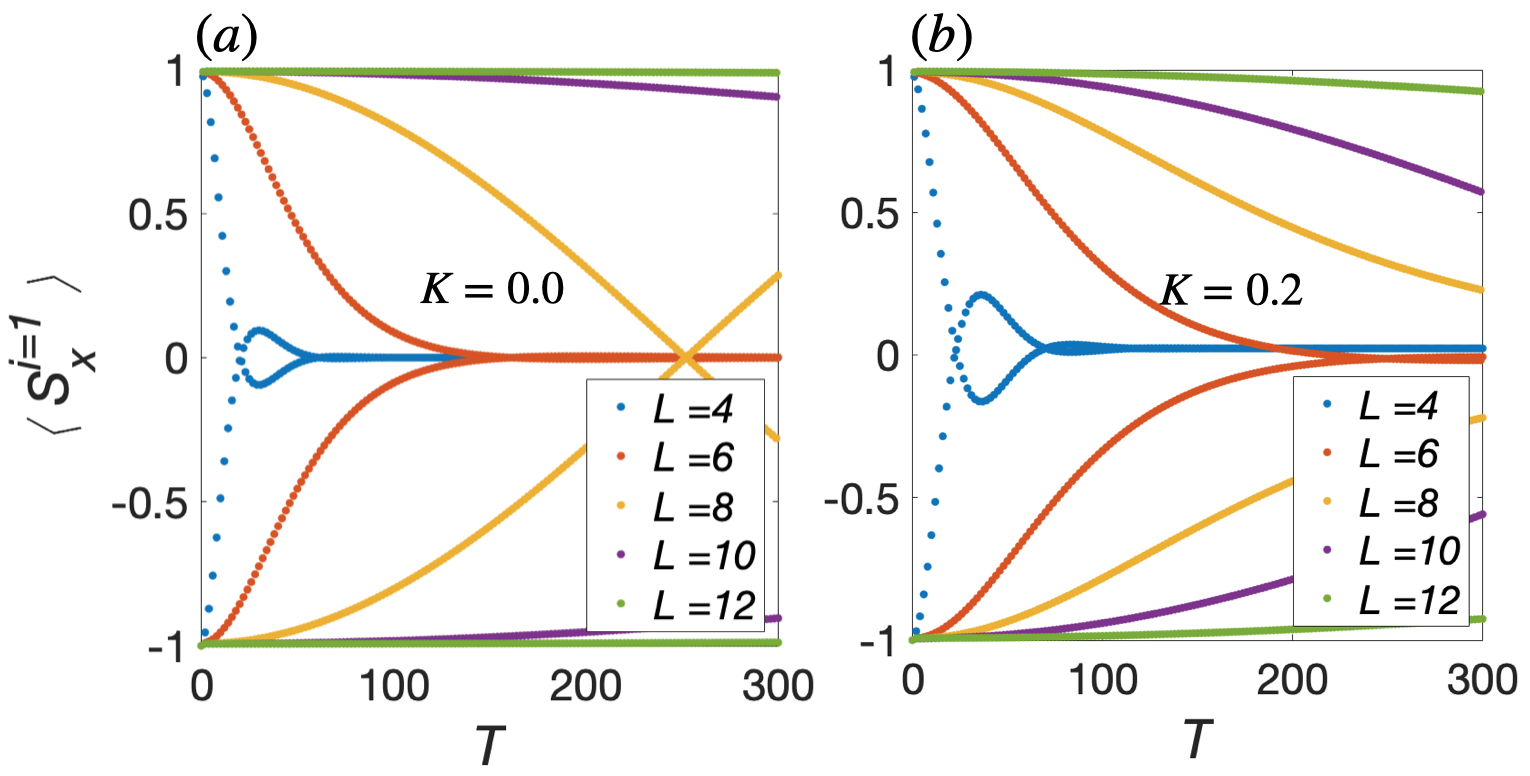}
\centering
\caption{Size dependence of $\langle S^{i=1}_x\rangle$ in the $(\pi)$ phase. $\alpha= 1.5$, $\beta_J = - 1.5 , \beta_h= 0.5 $ (in units of $\pi/4$). (a) Without integrability breaking $K = 0.0$ (b) With  integrability breaking $K = 0.2$ (in units of $\pi/4$).   The initial state is $\mid\downarrow,...\downarrow\rangle$.}
\label{spin}
\end{figure}

\section{Quantum quench in the spin language} 
\label{sec4} 
We have shown that zero modes and/or $\pi$ modes exist in the fermionic language if OBCs are imposed (see Fig.~\ref{edgemodes}), the same as in the unitary case \cite{Thakurathi2013floquet, yates2019almost}.  Different combinations of edge modes are expected to be associated with different evolutions of many-body states. 
In this section we summarize the results on the time evolution obtained directly in the spin language, starting from some random initial states (not close to Floquet eigenstates). We find four types of evolution corresponding to different phases in Fig.~\ref{pd}(b):

\begin{figure}[b]
\includegraphics[width=0.48\textwidth]{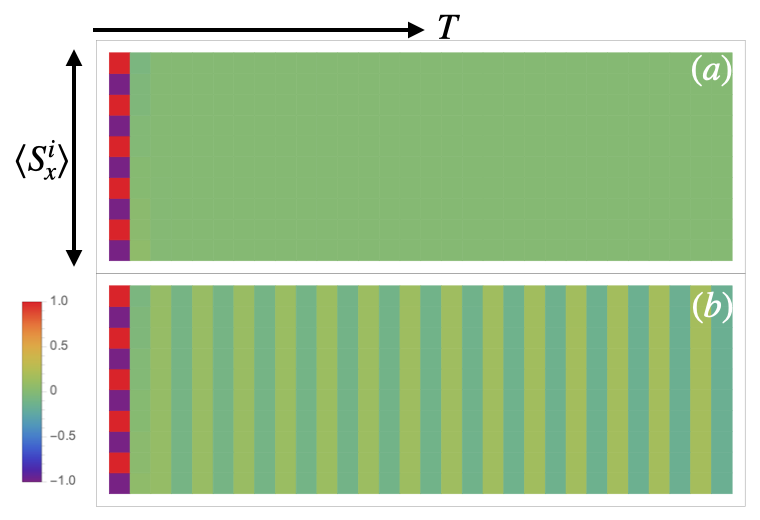}
\centering
\caption{Effect of the integrability breaking term $ K \sum_j X_j$ for a special initial (antiferromagnetic) state. (a) $K=0.0$; (b) $K =0.2$. $\alpha= 1.5$, $\beta_J = - 1.5 , \beta_h= 0.5 $ (in units of $\pi/4$).}
\label{phase5}
\end{figure} 

 \begin{figure}[t]
\includegraphics[width=0.48\textwidth]{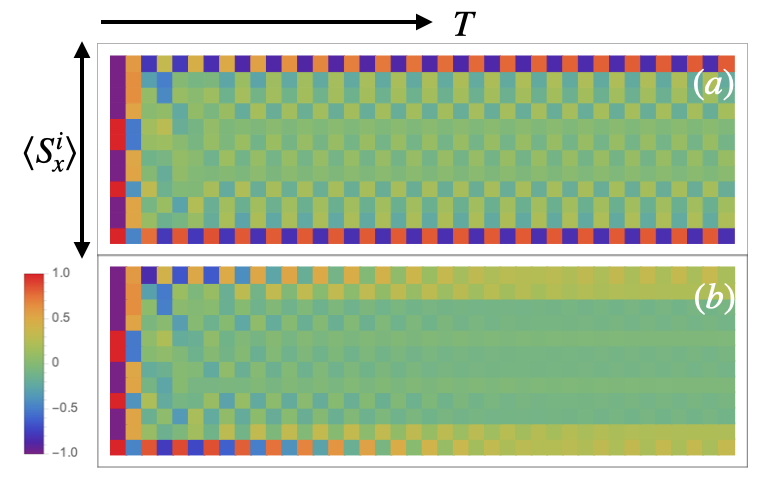}
\centering
\caption{Effect of the integrability-breaking term $ K \sum_j X_j$ in the $(0\pi)$ phase. (a) $K=0.0$; (b) $K =0.2$. $\alpha= 1.5$, $\beta_J =  -0.1 , \beta_h= 0.5 $  (in units of $\pi/4$).}
\label{phase6}
\end{figure} 

\begin{itemize}
\item In the trivial phase ($\oslash$), both $\langle S_x^i \rangle \to 0$ for all sites and $\langle S_x^1 S_x^N \rangle \to 0$, where $N$ is the total size, after a few drive periods [Fig.\ref{phase4}(a1)].  
\item There are two regions in the phase diagram with zero modes $(0)$. If $\beta_J <-\beta_h$, for a small system, $\langle S_x^i \rangle$ of the state rapidly drops to zero while $\langle S_x^1 S_x^N \rangle$ stays finite. For a small system, the final state has a large overlap with the Greenberger–Horne–Zeilinger state, $(\mid\uparrow\rangle^{\otimes L} + \mid\downarrow\rangle^{\otimes L})/\sqrt{2}$ (though it is very sensitive to integrability breaking terms).  Increasing the size slows down the initial decay of $\langle S_x^i \rangle$. In the thermodynamic limit, both $\langle S_x^i \rangle$ and $\langle S_x^1 S_x^N \rangle$ should stabilize to a finite value, meaning that the steady state breaks the spin-flip $\bbz_2$ symmetry [Fig.\ref{phase4}(b1)]. If $\beta_J >\beta_h$, the steady state is an antiferromagnet instead of a ferromagnet. 
\item In the phases with a $\pi$ mode $(\pi)$,  the situation is similar to the one with  the zero mode, except that the spins keep flipping, breaking the $\bbz_2$ time-translational symmetry [Fig.\ref{phase4}(c1)].  A typical evolution of $\langle S_x^{1} \rangle$ of different sizes is shown in Fig.~\ref{spin}(a). We expect that in the thermodynamic limit, the steady state is in a Floquet  (discrete time-crystal) phase. Moreover, if $\beta_J >\beta_h$, the steady state also breaks the $\bbz_2$ translational symmetry, corresponding to an antiferromagnet as opposed to the case $\beta_J <-\beta_h$ when it is a ferromagnet [Fig.\ref{phase4}(c1)].

\item In the phase with both zero modes and $\pi$ modes $(0\pi)$, the spins in the bulk $ \langle S_x^i \rangle$ and the spins near the edges behave differently as illustrated in Fig.~\ref{phase4}(d1).  We take the initial state as a product state of polarized spins. Bulk spin correlations $
\lim_{|i-j|\gg 1}\langle S_x^i S_x^j \rangle$ drop to zero, in contrast to the phases with only zero or $\pi$ modes. The steady states depend on $\beta_J$ and $\beta_h$. Thus, the initial behavior of different initial states can be different. In general, the oscillating pattern can be observed but edge spins have larger amplitudes.  We expect that the bulk spin amplitudes approach zero in the thermodynamic limit, while the edge oscillations persist.

\end{itemize}
In the above discussion, OBCs were used. The main features in the $(0)$ phase and the $(\pi)$ phase are basically unaffected if PBCs are used. This implies emergence of the long-range order in the bulk. However, the $(0\pi)$-phase depends crucially on the boundary conditions: Since this phase is evidenced solely by the edge spin oscillations, this signature disappears if we use PBCs. 


We next compare these different dynamics with those in the unitary case. To this end, we plot quench dynamics with different real parameters in Figs.\ref{phase4}(a2)-\ref{phase4}(d2). The four panels correspond to the same  combinations of edge modes as in the non-unitary case. In a clean system, we can see that the edge states are robust. These features can be attributed to the presence of (almost) strong edge modes \cite{fendley2016strong, Yates2020, yates2019almost}.  However, the bulk flows to the ``infinite-temperature" state very quickly (up to revivals due to the finite system size). In the magnetically ordered phases, $(0)$ and $(\pi)$, physically this can be understood in terms of domain-wall dynamics: domain walls can  move freely in the absence of disorder, quickly  destroying the bulk order; however, they cannot flip the edge spins, since that would lead to a change in the number of domain walls, and thus significantly change the quasienergy. 
The situation is very different from the system with disorder (MBL regime) \cite{Khemani2016phase}, where there are four different Floquet-MBL phases, each  characterized by a unique eigenstate order. Apparently, comparing the figures on the left and on the right in Fig.~\ref{phase4}, we find that the impact of imaginary parts of $J$ and $h$ is significant. In particular, the long-range bulk orders cannot be stabilized without the imaginary parts, i.e. measurements.

Since the TFIM is integrable, the steady states may depend on the initial states. Indeed, we see that if we start with some special initial states, say, with $\sum_i \langle S_x^i \rangle =0$, the quench dynamics can have a different behavior. One example in the $(\pi)$ phase is given in Fig.~\ref{phase5}(a) where the initial state is the antiferromagnetic state. This state has no overlap with the ferromagnetic $(\pi)$ steady state [see Fig. \ref{phase4}(b1)] and as a result becomes featureless  very quickly. However, if we add a small longitudinal field $K \sum_j X_j$ to break the integrability, the steady state reminiscent of Fig. \ref{phase4}(b1) above reemerges [Fig.\ref{phase5}(b)]. 

We also studied the effect  of the same integrability-breaking term $K \sum_j X_j$ on dynamics. The dependence of the evolution on the system size in the $(\pi)$ phase when $K \neq 0$ is shown in Fig.~\ref{spin}(b). As we can see, when we increase the system size, the decay rate drops rapidly to zero. A similar pattern in the $(0)$ phase is observed. Thus we expect these phases to be stable in the thermodynamic limit under the small perturbation ($0 < K \ll 1$).   As can be seen from Fig.\ref{phase6},  in the $(0\pi)$ phase, a small $K$ can polarize the spins, but the edge-bulk distinction survives for a long time.    This observation suggests that the influence of almost strong edge modes remains beyond the  unitary case discussed in Ref.  \cite{yates2019almost}. It is interesting to compare this behavior with the observation in the Google simulation in Ref.~\cite{Mi2022noise}, where they observed that the edge spins under the unitary $U_F$ evolution, in contrast to bulk spins, are resilient to integrability- and symmetry-breaking effects and dephasing effects such as low-frequency noise.  

Finally, note that $\alpha_J =\alpha_h$ is not necessary to obtain different dynamics regimes in the nonunitary case that we identified above. We also expect that the main features are robust against other non-Hermitian deformations.

\section{Conformal field theory at $J =h$}
\label{sec5}
\begin{figure}[b]
\includegraphics[width=0.42\textwidth]{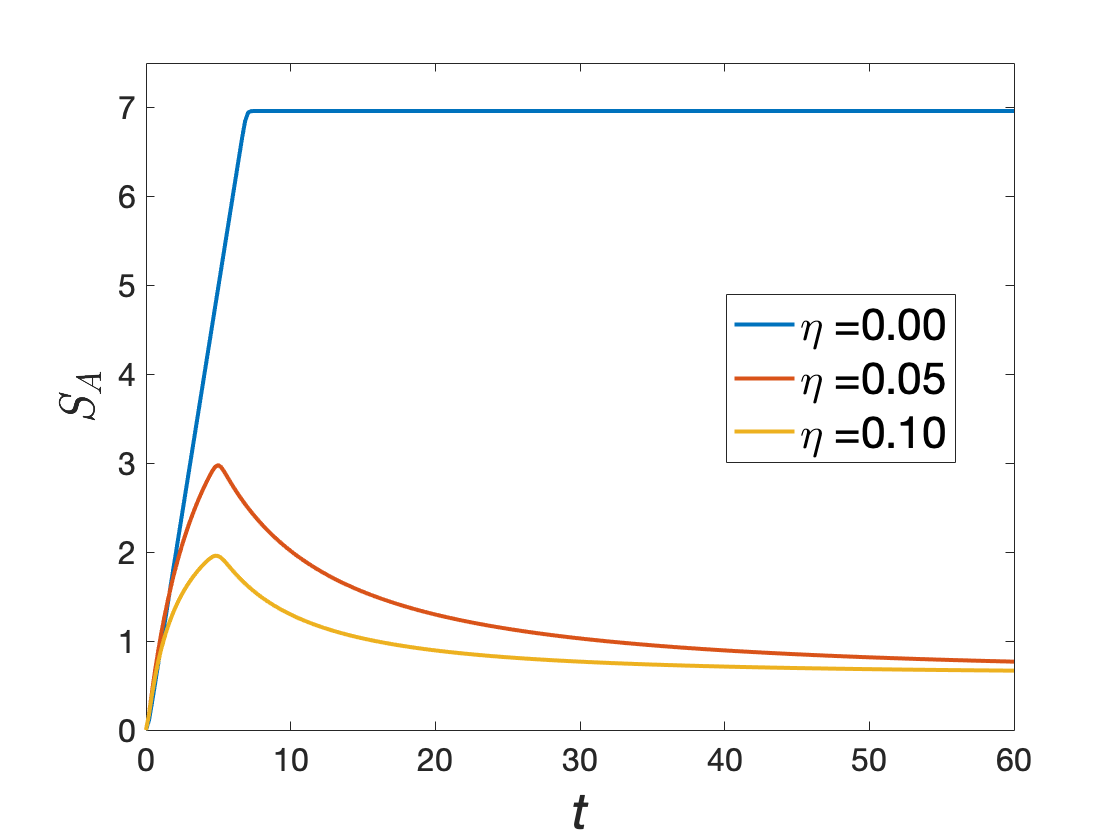}
\centering
\caption{Evolution of $S_A$ in a CFT [Eq.~(\ref{saa})] with different $\eta$. $c=1/2$, $\epsilon=0.185$, $l =L_A =10$.}
\label{cft1}
\end{figure}

So far, our focus has been on the area-law phases as well as volume-law critical lines in the phase diagram shown in Fig.~\ref{pd}(b). Now let us shift attention to the critical line with a logarithmic law at $J=h$. In the unitary case, i.e., $J$ and $h$ are real, if $|J| =|h| \to 0$, $U_F$ approaches that of the TFIM in the continuous time limit. The critical point is the celebrated Ising critical point described by the Ising CFT and the quench problem is well studied \cite{Calabrese2005Evolution, Calabrese2009entanglement}. It is interpretable using the quasiparticle picture \cite{Calabrese2005Evolution}. 
In the general case, the critical line (stabilized by disorder or interactions) may be called Floquet quantum criticality \cite{berdanier2018floquet}. In this section, we extend $J$ and $h$ to complex values and study the quench problem.

\begin{figure}[t]
\includegraphics[width=0.48\textwidth]{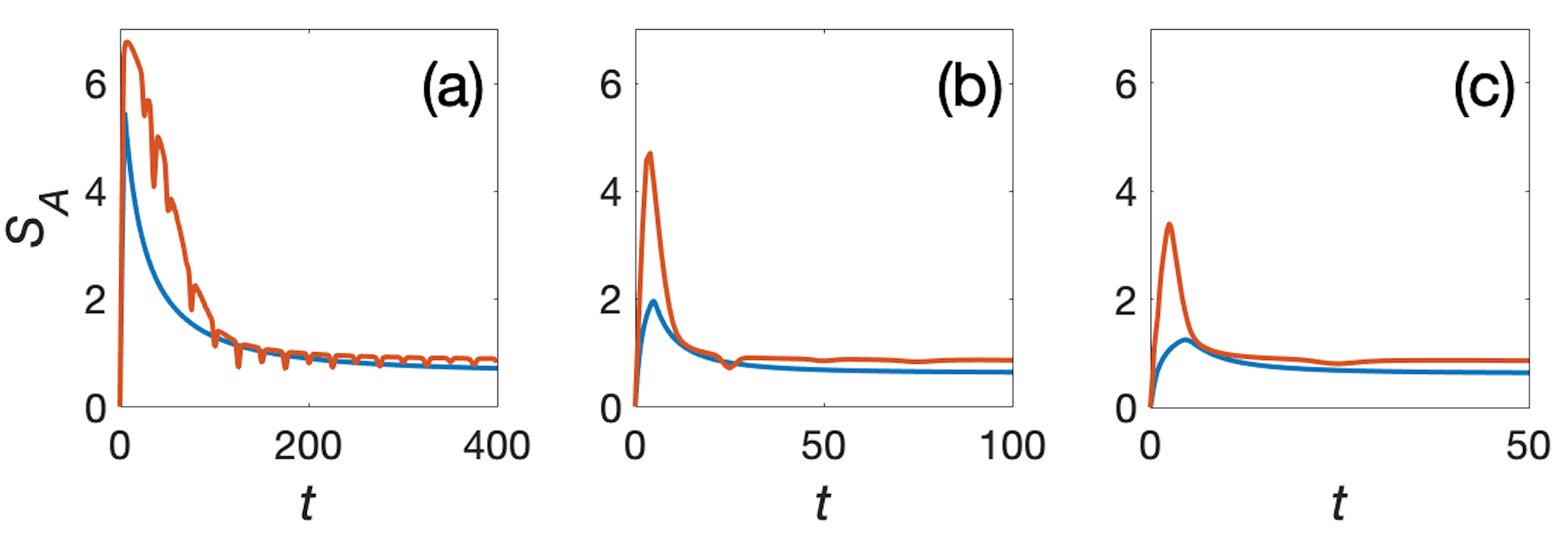}
\centering
\caption{Comparison of the evolution of $S_A$ of a CFT and the numerical result of the nonunitary TFIM.  The evolution in a CFT [Eq.~(\ref{saa})] with $c=1/2$, $\epsilon=0.185$, and $l= L_A =10$ (blue). The evolution under $U_F$ of the continuous complex TFIM [Eq.~(\ref{u_continuous})] with $J= h =1- i \eta $, $L_A =10$ and $L_A/L=1/10$ (red). (a) $\eta =0.01$, (b) $\eta =0.10$, (c) $\eta =0.20$. The oscillations are due to the finite-size effects. }
\label{cft3}
\end{figure}

\subsection{Continuous-time limit}
When $J =h$ and $|J| =|h|\to 0$, the qualitative behavior of the system is captured by the continuous time limit where the dispersion is given by $\pm 4 J |\sin(k/2)|$ [Eq.~(\ref{disp})]. For complex $J$ and $h$, the dispersion is rotated to the complex plane in general. If we assume that the formalism developed by Calabrese and Cardy \cite{Calabrese2005Evolution} can be generalized to this case,  the quench problem can equivalently be described by $\exp( -i (1-i\eta) t H_{\text{CFT}})$, where $\eta \ge 0$ quantifies the rotation. For more information, see the Appendix. Since $\exp( -i (1-i\eta) t H_{\text{CFT}}) = \exp( -i t H_{\text{CFT}}) \exp( -\eta t H_{\text{CFT}})$, the initial state first evolves by  $\exp( -\eta t H_{\text{CFT}})$ then by the unitary operator $\exp( -i t H_{\text{CFT}})$. Since $\exp( -\eta t H_{\text{CFT}})$ evolves the initial state to the ground state of $H_{\text{CFT}}$ asymptotically, the long-time evolution approaches that of the ground state. 

As discussed in  the Appendix, the von Neumann entanglement entropy is
\be 
S_A(t) \equiv  \left[- \frac{\partial}{\partial n} \tr \rho_A^n (t)   \right]\vert_{n=1}.  
\label{saa}
\ee
with 
\begin{align}
& \tr \rho_A^n (t)  \simeq   \nonumber \\
&    \left( \frac{\pi}{2\tau_0}\right)^{2d_n} \left(\frac{\cosh(\pi l/2\tau_0) + \cosh (\pi t/\tau_0)}{8 \sinh(\pi l/4\tau_0)^2 \cosh^2 (\pi t/2\tau_0)} \right)^{d_n}  
\end{align}
 where 
$d_n = (c/12)(n -1/n)$, $c$ being the central charge, and $\tau_0 = \epsilon + \eta t$, $\epsilon$ being a regularization constant. In the derivation of the above equations, we have  assumed $t, l \gg \tau_0$.

Typical evolutions of (normalized) von Neumann entanglement entropy $S_A$ are depicted in Fig.~\ref{cft1}. $\epsilon =0.185$ is chosen such that if $\eta =0$,  $S_A$ for $l = L_A$ saturates at the same value $S_A \sim \pi c l/12\epsilon $ of the Ising CFT ($c =1/2$). When $\eta$ becomes finite, $S_A$ keeps increasing until $t \sim l/2$, then it starts to drop. In a closed system, the regularization $\epsilon$ affects both the initial growth rate and the saturation value of entanglement entropy. With dissipation $\eta$, $\epsilon$ becomes less important after some time: $ t > \epsilon/\eta$. However, $\epsilon$ determines not only the initial maximum of $S_A$ but also the ground state entanglement entropy. The long time limit depends on $\epsilon$ but not $\eta$ because as we mentioned the state approaches the ground state. The larger $\eta$ is, the faster the decaying rate is. In general, $S_A$ can have a volume law before gradually approaching a logarithmic law at long times.  The evolutions of $S_A$ of the TFIM with continuous time and different $\eta$ [Eq.~(\ref{u_continuous})] are compared with those predicted by Eq.~(\ref{saa}) in Fig.~\ref{cft3}. We see that they match qualitatively.

Note that even thought $H_{\text{CFT}}$ is the critical Ising Hamiltonian in our main discussion, the formalism of Calabrese and Cardy \cite{Calabrese2005Evolution} applies to general critical $H_{\text{CFT}}$. Thus, we expect the simple qualitative prediction from the CFT to be much more general. Also, as mentioned in Sec.~\ref{alpha1}, the pattern of the evolution of $S_A$ i.e., increasing first then decaying, is shared by the area law phases. Similar features were observed in dissipative systems (see e.g., Refs. \cite{Alba2021spreading, Carollo2022dissipative}).

\begin{figure}[tb]
\includegraphics[width=0.48\textwidth]{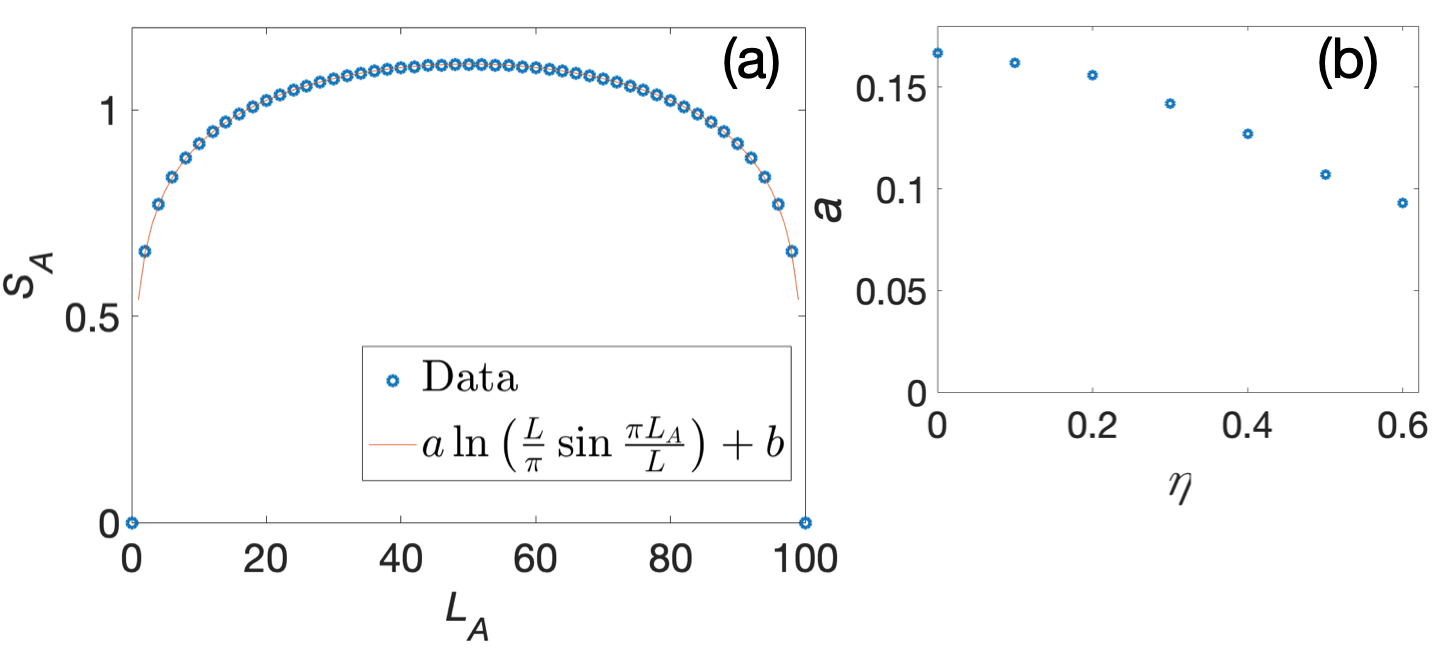}
\centering
\caption{(a)
Subsystem dependence of  $S_A$ in a total system with $L=100$ with fit in Eq.~(\ref{fit}) using  $J= h = 0.2 -0.1 i$ (in units of $\pi/4$). $a \approx 0.165$ and $b \approx 0.54$.
(b) Fitting values of $a$ as a function of $\eta$. $J= h = \eta i$ (in units of $\pi/4$).}
\label{cft6}
\end{figure}

\subsection{Floquet criticality}
Since there is no phase transition until $\alpha_J = \alpha_h = \pi/4$, even though the preceding discussion was focused on the continuous time limit, the evolution of entanglement entropy in Fig.~\ref{cft1} should qualitatively apply more broadly (i.e., for finite $J =h$). Once we move away from the continuous time limit, the Floquet criticality \cite{berdanier2018floquet} is no longer the canonical $c=1/2$ Ising CFT. In Fig.~\ref{cft6}(a), we fix the total system size $L =100$ and plot $S_A$ against different subsystem sizes $L_A$. Then the numerical data are fit by \be a \ln \left(\frac{L}{\pi} \sin \frac{\pi L_A}{L} \right) +b\label{fit}
\ee to extract the central charge \cite{Calabrese2009entanglement, Lavasani2021measurement}. 
In the continuous time limit, $a = 1/6 =c/3$ since $c=1/2$. In the Floquet setting, both $a$ and $b$ depend on $J=h$. The fitting in  Eq.~(\ref{fit}) applies as long as $\alpha =\alpha_J = \alpha_h$ is away from $\pi/4$, the critical line where the quench resembles  that in the unitary case with an extensive number of real modes. For example, if $J= h = 0.2 -0.1 i$ (in units of $\pi/4$), we find $a \approx 0.165$ and $b \approx 0.54$.  In general, the fit $a$ depends on imaginary part $\beta$ but not the real part $\alpha$. Larger $\beta$ leads to smaller $a$. In Fig.~\ref{cft6}(b), we set $\alpha =0$ and $\beta = \eta$, i.e. $J= h = \eta i$, and plot $a$ for several values of $\eta$. We see that it decreases as a function of $\eta$.

\section{Discussion} 
\label{sec6}
In this work, we have studied the nonunitary Floquet TFIM (kicked Ising model) with complex couplings and transverse fields. We analyzed the spectrum of the Floquet Hamiltonian in the fermionic language using the Jordan-Wigner transformation with both PBCs and OBCs, and used it to map out the phase diagram. For PBCs, we found that the spectrum may contain no real modes, a few real modes, or a finite density of real modes, the latter enabled by  pseudo-Hermiticity of the Hamiltonian for special values of parameters. For OBCs,  we found that real zero- and/or $\pi$ edge modes can exist in different phases. We presented the numerical result for the evolution of subsystem entanglement entropy after a quantum quench. In general, the entanglement entropy  increases initially,  then begins to drop, and eventually  saturates to some steady-state value. The three just mentioned spectral cases lead to an area law, a logarithmic law, and a volume law of entanglement entropy in the steady state, respectively. The scaling behavior can be interpreted in terms of real-energy non-Hermitian quasiparticles being responsible for establishing entanglement (the initial entanglement peak can be attributed to complex-energy quasiparticles, which can lead to enhanced entanglement on the time scales of their lifetime).
A quantized TEE exists if a zero mode or $\pi$ mode exists.  We also studied the quench dynamics of an open spin chain from a typical initial state and identified four types of dynamical behaviors corresponding to different edge mode configurations. Compared to the clean unitary case, which generally lacks bulk long range order, bulk order is stabilized by nonzero imaginary parts of $J$ and $h$, i.e. measurements. Finally, we considered the case when $J =h$ and compared the numerical results in the continuous time limit with the analytical result of a CFT by extending Calabrese and Cardy's formalism to complex time. They match qualitatively. Also, the effective central charge of the Floquet criticality was found to depend on $J =h$. 
 
There are many promising future directions. First, there are many observables that can be used to study the nonunitary Floquet TFIM.
In this work, we only computed entanglement entropy. As a matter of fact, we found mutual information has a similar behavior. It is also possible to compute other quantities such as entanglement negativity \cite{vidal2002}. In particular, the general CFT formalism can also be generalized properly to give us some insights into the evolution and the scaling of these quantities. In our work, we only discussed the non-Hermitian generalization of the quasiparticle picture at the spectral level; a quantitative check on the entanglement entropy growth as in Ref. \cite{alba2017, Alba2021spreading, Carollo2022dissipative} should be done. It is also interesting to see if more general topological quantities can be constructed to distinguish all four cases [so far we found that the $(0,\pi)$ phase is not clearly detected by TEE, in contrast to  $(0)$ and $(\pi)$  phases].  In addition, the relation to the skin effects in the context of Floquet non-Hermitian topological phases \cite{banerjee2023non, zhou2023non} should be explored \cite{bergholtz2021}.   Also, we have focused on 1D spin chains. Some of our discussion for the 1D case can be easily generalized to higher dimensions.   

Second, we have used non-Hermitian Hamiltonians as our starting point to study the effect of measurements by complexifying the coefficients.
Complex $J$ and $h$ correspond to post-selecting no-jump trajectories,  with the jump  operators $L_i = (1\pm X_i X_{i+1})/2$ and $L_i = (1\pm Z_i)/2$, respectively. There are many other types of measurements. We hope that the results of the simplified system can at least yield some insights into dynamical systems described by the stochastic Schrodinger equation where there are few or no post-selections \cite{kells2021topological}.  How more general continuously monitored systems behave deserves to  be studied more closely, see, e.g., Ref.~\cite{Cao2019Entanglement, Alberton2021Entanglement, Carollo2022dissipative}. Whether different spin dynamics of a dissipative Floquet TFIM survive under a quench without post-selection should be explored \cite{Sierant2022dissipativefloquet, kells2021topological, Mi2022noise, mi2023stable}.

Third, 
there are many ways to extend the non-Hermitian Hamiltonian.  Other physical effective Hamiltonians can be studied in a similar way. In this work we only briefly discussed the effect of  the integrability-breaking term $\sum h'_i X_i$ in the Hamiltonian.  We can also include terms like $J_z \sum  Z_i Z_{i+1}$ which after the Jordan-Wigner transformation is mapped to a density-density interaction \cite{Khemani2016phase}. Introducing long range interactions is another direction with possible experimental relevance \cite{Sierant2022dissipativefloquet}. We know that introducing spatial disorder in the unitary case into such a system with a small $J_z$  leads to MBL and stabilizes the Floquet phases \cite{Khemani2016phase, moessner2017equilibration}. How these systems respond to complexification of the coefficients is an interesting open question.

Lastly, several more general ideas discussed in this work deserve  further investigations. 
For example, the partially real spectrum on some critical lines implies a spontaneous breaking of the antilinear symmetry associated with a diagonalizable pseudo-Hermitian Hamiltonian \cite{mostafazadeh2002pseudo} as a function of momentum $k$  (rather than some external control parameter). The effect of the existence of such an exceptional point in the $k$ space on entanglement and purification transition should be further explored \cite{Gopalakrishnan2021Entanglement}. 
As another example, we noticed that positive or negative decay rates effectively lead to projective measurements in the momentum space. A general question is whether engineered dissipation can be used to construct interesting phases like topological phases \cite{kraus2008preparation,verstraete2009quantum, harrington2022engineered}.

We thank Rahul Sahay for discussions. This material is based upon work supported by Laboratory Directed Research and Development (LDRD) funding from Argonne National Laboratory, provided by the Director, Office of Science, of the U.S. Department of Energy under Contract No. DE-AC02-06CH11357.

\bibliographystyle{jhep}
\bibliography{TFIM}  

\appendix

\section{Conformal field theory with complex time}
\label{appCFT}
The evolution of entanglement entropy of a unitary critical system after a quantum quench is discussed in Refs.~\cite{Calabrese2005Evolution, Calabrese2009entanglement}. We discuss a minor generalization to a special critical non-Hermitian system by rotating the real time to the complex plane.

If $H$ is the Hamiltonian of a CFT, then the non-Hermitian Hamiltonian $(1-i \eta) H$ with $\eta >0$ may be regarded as a critical Hamiltonian of the non-Hermitian system since the zero mode remains intact while other modes become complex. In our context, when $J = h= \alpha (1-i \eta)$ and $|J|\ll 1$, the effective Hamiltonian is approximately described by that of a critical TFIM but with complex couplings. We assume that the CFT formalism is applicable to $(1-i \eta) H$ and show that the CFT formalism yields results that are compatible with those in Ref.~\cite{Carollo2022dissipative} qualitatively. 

Suppose the system starts with the initial state $|\psi_0\rangle$ at $t=0$ and then starts to evolve under $(1-i \eta) H$. We first compute the density matrix  
\begin{align}
& \langle \psi (x, t)| \rho_0| \psi'(x', t)\rangle\nonumber \\
=& Z^{-1} \langle \psi (x) | e^{i t (1+i \eta) H -\epsilon H} \rho_0  e^{-i t (1-i \eta) H -\epsilon H} |\psi'(x')\rangle \nonumber \\
=& Z^{-1}\langle \psi (x) | e^{- (t \eta +\epsilon - i t) H} \rho_0  e^{ -( t\eta+\epsilon +i t ) H} |\psi'(x')\rangle,
\end{align}
where $ \rho_0 =|\psi_0\rangle\langle \psi_0|$,  $Z = \tr [ \rho_0 e^{-2(\epsilon + t \eta) H}]$, and a  term $\epsilon H$ has been added to make the quantity convergent.

Let us write $\tau_1 = \epsilon + t \eta + it$ and $\tau_2 = \epsilon + t \eta - it$, then the density matrix $\rho$ can be represented as two path integrals: the first one starts with $\psi'(x', \tau)$ at $\tau =- \tau_1$ and ends with  $\psi_0(x)$ at $\tau =0$, and  the second one starts with $\psi_0(x)$ at $\tau =0$ and terminates with $\psi(x, \tau)$ at $\tau = \tau_2$. Divide the system (at $\tau =0$) into region $A$ and its complement $B$, then the reduced density matrix $\rho_A$ can be obtained by gluing $\psi_0(x)$ of the first integral with that of the second one at $\tau =0$ for $x$ inside region $B$. $\tr \rho_A^n$ can be obtained by cyclic gluing:
\be 
\tr \rho_A^n = Z_n(A)/Z^n,
\ee
where $Z_n(A)$ is the path integral on an $n$-sheet surface. Taking $n=1$ and $A$ to be the entire system, we have $Z_1 = Z$ and $\tr \rho =1$. The Renyi entropies are given by 
\be
S_A^{(n)} = \frac{1}{1-n} \tr \rho_A^n,
\ee
and the von Neumann entanglement entropy by 
\be
S_A = - \frac{\partial}{\partial n} \tr \rho_A^n\vert_{n=1}. 
\label{SA}
\ee
To compute $\tr \rho_A^n$, we make use of the conformal invariance and parametrize the strip with width $\tau = \tau_1+\tau_2$ by the upper half-plane of the complex plane via the conformal map $w = (2 \tau_0/\pi) \log z$ with $\tau_0 = \epsilon + \eta t$. If the total length of  $A$ is $l$ and the original end points of $A$ sit at $-l/2$ and $l/2$, the images of these two branch points are $z_1 = \exp( -\pi l/4\tau_0 )  \exp( i \pi \tau_1 /2\tau_0) $ and $z_2 = \exp(\pi l/4\tau_0 )  \exp( i \pi \tau_1 /2\tau_0) $, respectively. Following Ref.~\cite{Calabrese2005Evolution}, $Z_n(A)/Z^n$ can be computed from the correlation function of the two branch point twist fields with scaling dimension $d_n = (c/12)(n -1/n)$:
\begin{align}
& \tr \rho_A^n (t)  \simeq \tilde{\cal{F}}_n(x) \\
& \times   c_n \left( \frac{\pi}{2\tau_0}\right)^{2d_n} \left(\frac{\cosh(\pi l/2\tau_0) + \cosh (\pi t/\tau_0)}{8 \sinh(\pi l/4\tau_0)^2 \cosh^2 (\pi t/2\tau_0)} \right)^{d_n}  \nonumber  
\end{align}
where $\tilde{\cal{F}}_n(x)$ is a function that depends on the model and the boundary condition, $x$ being the four-point ratio, and $c_n$ are constants that cannot be determined with this method but $c_1 =1$.  If we are interested in asymptotic behaviors, e.g. $t, l \gg \tau_0$,  $\tilde{F}_n(x)$ is irrelevant and we can put it to be 1. After some algebraic manipulations, we obtain
\be
\tr \rho_A^n(t) \simeq c_n \left(\frac{\pi}{2\tau_0} \right)^{2d_n} \left( \frac{e^{\pi l/2\tau_0} + e^{\pi t/\tau_0}}{e^{\pi l/2\tau_0}  e^{\pi t/\tau_0}} \right)^{d_n}.
\label{eqrhon1}
\ee  
Note that since $\tau_0 = \epsilon + \eta t$,  for $t, l \gg \tau_0$ to be satisfied, we require that $\eta\ll 1$ and $ t \ll l/\eta$.  We can normalize the entanglement entropy by subtracting $S_A(0)$ from $S_A(t)$.
Using the approximation in Eq.~\ref{eqrhon1}, we find that normalized entanglement entropy is given by
\be
S_A(t) \simeq \frac{c}{3} \ln \left(\frac{ \epsilon + \eta t}{\epsilon }\right) +\begin{cases} \frac{\pi c t}{6(\epsilon + \eta t)} \quad & \epsilon\ll t<l/2 \\  \frac{\pi c l}{12(\epsilon + \eta t)} \quad &  l/2 < t\ll l/\eta \end{cases}.
\ee
If we take $ l/2 < t< l/\eta$ and $t\to \infty$, $S_A(t)$ will saturate. 

The derivation above can be generalized to Renyi entropies straightforwardly. In fact, Renyi entropies have similar behaviors as the von Neumann entanglement entropy.

\end{document}